\documentclass[%
 reprint,
 amsmath,
 amssymb,
 aps,
]{revtex4-2}

\usepackage[dvipdfmx]{graphicx}
\usepackage{dcolumn}
\usepackage{bm}
\usepackage{color}

\begin{document}

\preprint{APS/123-QED}

\title{Contribution of directedness in graph spectra}

\author{Masaki Ochi}
\email{ochi@iis.u-tokyo.ac.jp}
 \affiliation{Department of  Physics, The  University of Tokyo,\\
 5-1-5 Kashiwanoha, Kashiwa, Chiba, 277-8574, Japan}
 
\author{Tatsuro Kawamoto}%
 \email{kawamoto.tatsuro@aist.go.jp}
\affiliation{Artificial Intelligence Research Center, \\
  National Institute of Advanced Industrial Science and Technology, 2-3-26 Aomi, Koto-ku, 
  Tokyo, 135-0064, Japan }

\date{\today}

\begin{abstract}
In graph analyses, directed edges are often approximated to undirected ones so that the adjacency matrices may be symmetric. 
However, such a simplification has not been thoroughly verified. 
In this study, we investigate how directedness affects the graph spectra by introducing random directization, which is an opposite operation of neglecting edge directions. 
We analytically reveal that uniformly random directization typically conserves the relative spectral structure of the adjacency matrix in the perturbative regime.
The result of random directization implies that the spectrum of the adjacency matrix can be conserved after the directedness is ignored.

\end{abstract}

\maketitle

\section{\label{section-1}Introduction}

Many real-world datasets are represented by directed graphs. 
In social networks, the follower-followee relationship defines a directed edge \cite{mcauley2012learning}. 
In nervous systems, a signal transduction between neurons occurs only in one direction \cite{watts1998collective}. 
In this way, the directedness renders the relationship between a pair of vertices asymmetric and characterizes many properties on graphs, such as the diffusion \cite{newman2002email} and reciprocity \cite{garlaschelli2004patterns}.
Nevertheless, in the studies of complex networks, the edge directions in graphs are often ignored, and directed graphs are converted to the undirected counterparts. 
Here, we refer to such simplification as \textit{undirectization}.
While undirectization may affect the result of an analysis only negligibly, it can be critical in some cases. 
In this study, we investigate the importance of the directedness in graph analyses by focusing on the change in graph spectra using the matrix perturbation theory. 

Graphs are typically represented by matrices, such as adjacency matrices, combinatorial and normalized Laplacians \cite{von2007tutorial},  and non-backtracking matrices \cite{krzakala2013spectral}. 
The associated graph spectra offer important tools for capturing the global properties of graphs, such as module structures \cite{newman2006finding} and network centralities \cite{bonacich1987power}. 
For example, in spectral clustering for undirected graphs, the number of clusters is determined by the number of eigenvalues that are isolated from the (asymptotically) continuous spectral band. 
The eigenvectors corresponding to these isolated eigenvalues provide partitioning of a graph \cite{fortunato2010community,von2007tutorial}. 
There have been a number of studies on the spectral properties for variety of matrices in the context of statistical physics and random matrix theory \cite{rogers2008cavity,kuhn2008spectra,nadakuditi2012graph}.
In spectral graph theory, several bounds for the largest and second-largest eigenvalue for adjacency matrices and Laplacians have been studied \cite{chung2004spectra,das2004some,nilli1991second}.

\begin{figure}[t!]
    \begin{tabular}{c}
    \begin{minipage}[b]{0.95\linewidth}
    \centering
    \includegraphics[width=\linewidth]{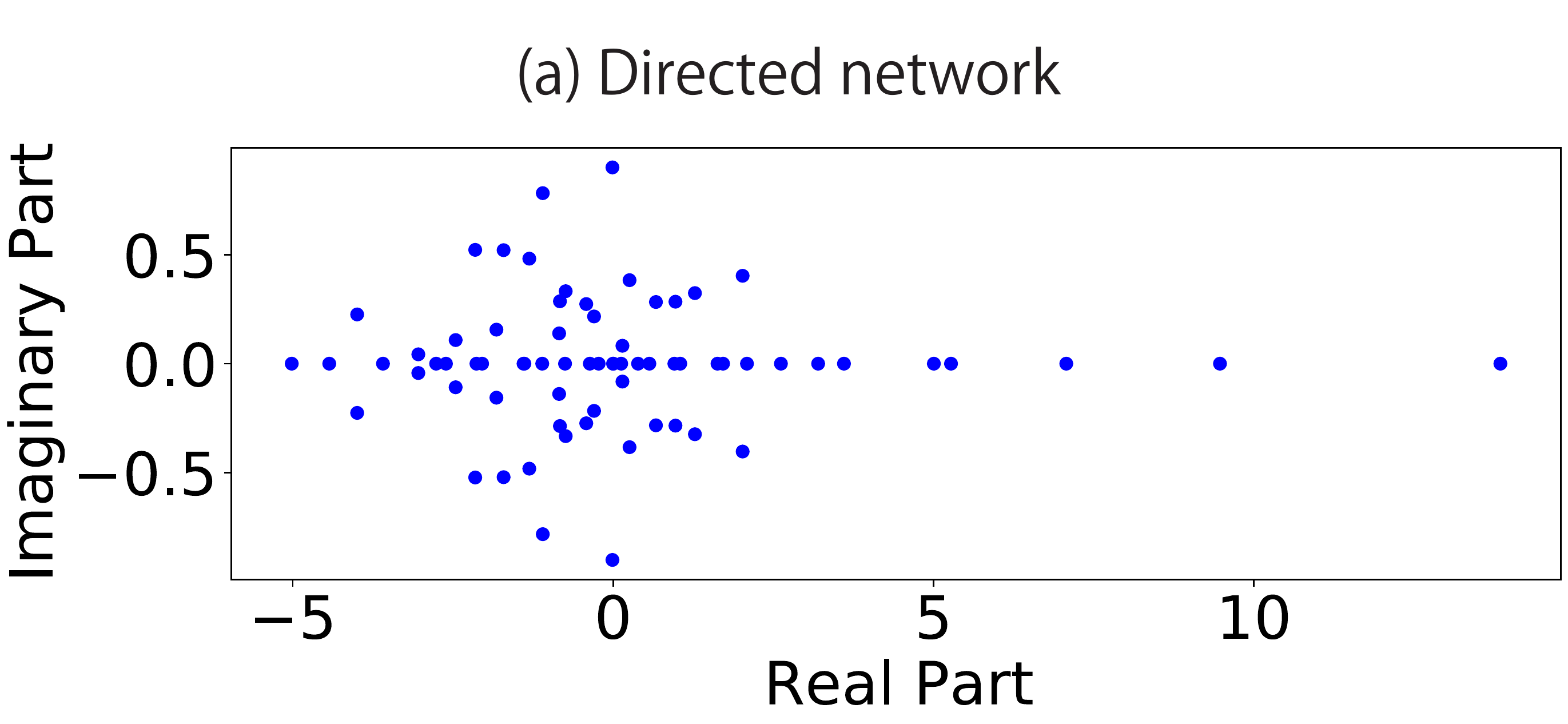}
    \end{minipage}\\
    \begin{minipage}[b]{0.95\linewidth}
    \centering
    \includegraphics[width=\linewidth]{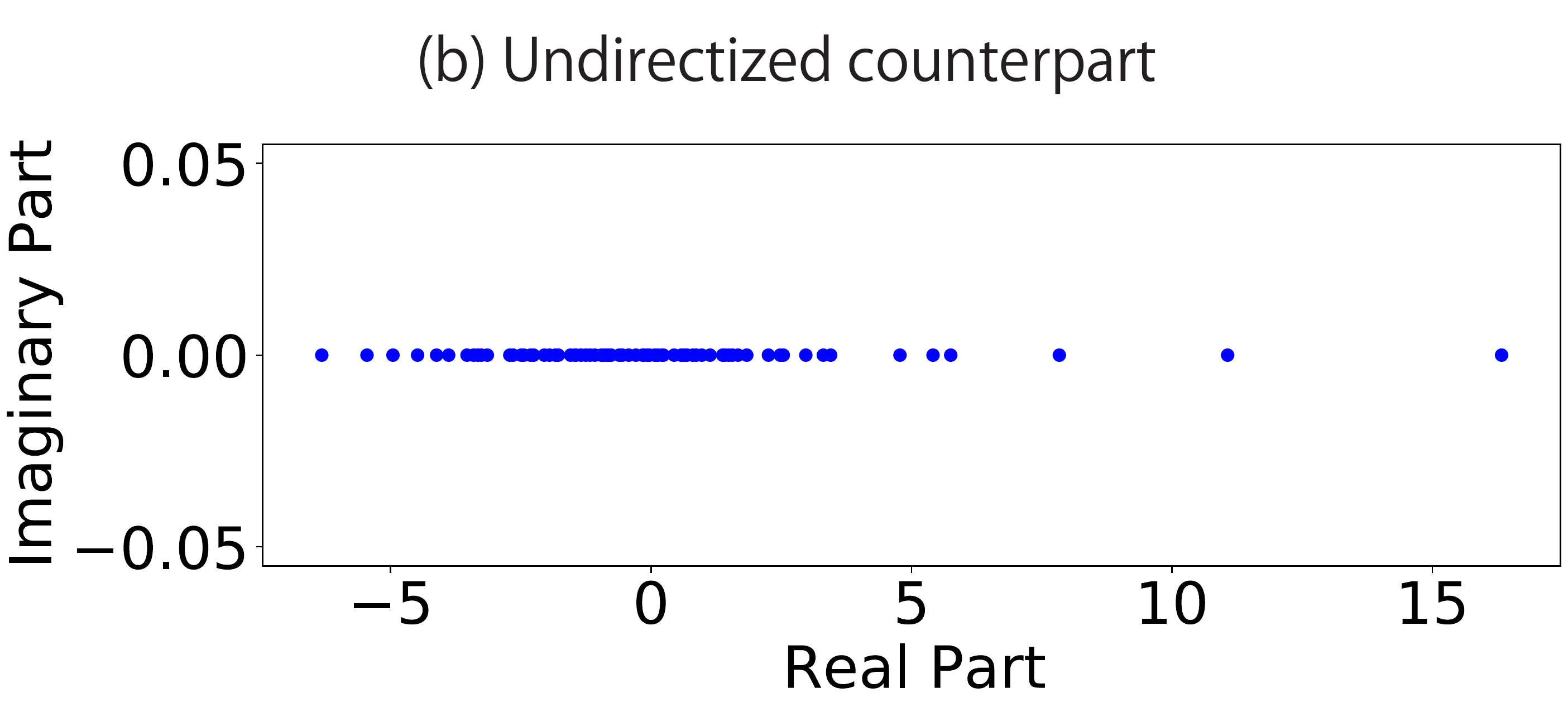}
    \end{minipage}
    \end{tabular}
    \caption{Adjacency matrix spectra for the macaque-cortex network \cite{young1993organization}: (a) original network with directed edges and (b)  undirectized network in which every directed edge is converted to an undirected edge.}
    \label{fig-sec1-1}
\end{figure}

Spectral methods for undirected graphs and directed graphs are not completely analogous to each other. 
Because the matrices are asymmetric for directed graphs, their spectra contain eigenvalues with non-zero imaginary parts.
This is partly a reason that motivates researchers and practitioners to ignore edge directions.
To formulate spectral methods for directed graphs, several graph Laplacians for the spectral clustering of directed graphs have been proposed in the literature \cite{chung2005laplacians,zhou2005learning,li2012digraph,malliaros2013clustering,rohe2016co,yoshida2016nonlinear}. 
Although the choice of Laplacians is an important issue, herein, we focus only on adjacency matrices of directed and undirected graphs.
There are several researches on the spectra of directed graphs.
For example, in spectral graph theory, bounds of the spectral radius of adjacency matrices for directed graphs have been studied \cite{brualdi2010spectra,gudino2010lower}.
In random graph theory and statistical physics, spectral densities of adjacency matrices for random directed graphs have been investigated \cite{sommers1988spectrum,rogers2009cavity}.
In contrast, we consider typical spectra of directed graphs based on its undirected counterpart.

Figure~\ref{fig-sec1-1} (a) shows the eigenvalue distribution of the adjacency matrix of a macaque-cortex network \cite{young1993organization}, which is partially directed; Fig.~\ref{fig-sec1-1}(b) presents the undirectized counterpart. 
We note that all eigenvalues are projected onto the real axis, and the scale along the real part is slightly larger in the undirectized graph.
Despite these differences, the relative distances among the five largest eigenvalues along the real axis are almost unchanged between Figs.~\ref{fig-sec1-1}(a) and \ref{fig-sec1-1}(b).

The last observation in this example motivates us to theoretically investigate the relationship between the spectral structures of directed graphs and their undirectized counterparts.
To this end, we introduce \textit{random directization} as an opposite operation of ignoring edge directions.
That is, we consider an undirected graph as the original graph and randomly make undirected edges directed. 
We consider the typical variation of eigenvalues and eigenvectors under random directization.
When the fraction of directized edges is sufficiently small compared to the total number of edges, the resulting adjacency matrix can be considered as a perturbed matrix of the original one.
We apply the matrix perturbation theory to analytically evaluate variations of eigenvalues and eigenvectors after directization.
As shown below, an important prediction of the perturbation theory is that the relative spectral structure along the real axis of the adjacency matrix is approximately conserved when the edges are directized uniformly randomly. 
This conversely explains our observation in Fig.~\ref{fig-sec1-1} on undirectization.

There have been many works on perturbative analysis for undirected graph spectra.
Let $\bm{A}$ and $\bm{V}$ be real-symmetric $n\times n$ matrices, and we perturb $\bm{A}$ by adding $\bm{V}$.
Let $\lambda_i$, $\bm{v}_i$ respectively denote the $i$th eigenvalue and eigenvector of $\bm{A}$, and $\tilde{\lambda}_i$, $\tilde{\bm{v}}_i$ respectively denote the $i$th eigenvalue and eigenvector of $\bm{A}+\bm{V}$.
The bound for the variation of eigenvalues through this perturbation is known as the Weyl's theorem \cite{Weyl1912Das}:
\begin{align}
\tilde{\lambda}_i -\lambda_i \leq ||\bm{V}||,
\end{align}
where $||\bm{V}||$ represents the spectral norm of $\bm{V}$.
As for the variation of eigenvectors, the Davis-Kahan theorem \cite{davis1970rotation} explains how the eigenvector can change with the same perturbation: the angle between $\bm{v}_i$ and $\tilde{\bm{v}}_i$ is bounded as 
\begin{equation}
\sin \angle(\bm{v}_i,\tilde{\bm{v}}_i)\leq\frac{2||\bm{V}||}{\Delta_i}, \quad \Delta_i = \min\{|\tilde{\lambda_j}-\lambda_i|:  j\neq i\},
\end{equation}
where the sine of the angle between two vectors is defined by $\sin \angle (\bm{v}, \bm{w})=\sqrt{1-\left(\bm{v}\cdot\bm{w}/|\bm{v}||\bm{w}|\right)^2}$.
In addition, many variants of the Weyl's and Davis-Kahan theorems, such as the one convenient for application in statistical contexts \cite{yu2015useful} and the ones which assume low-rank matrices for $\bm{A}$ and random matrices for $\bm{V}$ \cite{fan2018eigenvector,eldridge2018unperturbed}.
Sarkar \textit{et al}.\ \cite{sarkar2016eigenvector} extended the theorem in order to algorithmically estimate the number of modules in undirected graphs.
Karrer and Newman \cite{karrer2011stochastic} experimentally investigated the effect of adding edges to undirected graphs on their spectra.
Note that in these studies, both the unperturbed and perturbed graphs were undirected, whereas in our study, we consider directizing perturbation.

The rest of the paper is organized as follows.
In Sec.~\ref{section-2}, we formally define random directization and conduct a perturbative analysis. 
Then, we evaluate the variation in the spectra under undirectization in Sec.~\ref{section-3}.
Finally, Sec.~\ref{section-4} is devoted to a summary and discussions.
The symbols used in this paper are listed in Appendix~\ref{sec:app-notation}.

\section{\label{section-2}Random directization}

\begin{figure}[t!]
    \includegraphics[width=\linewidth]{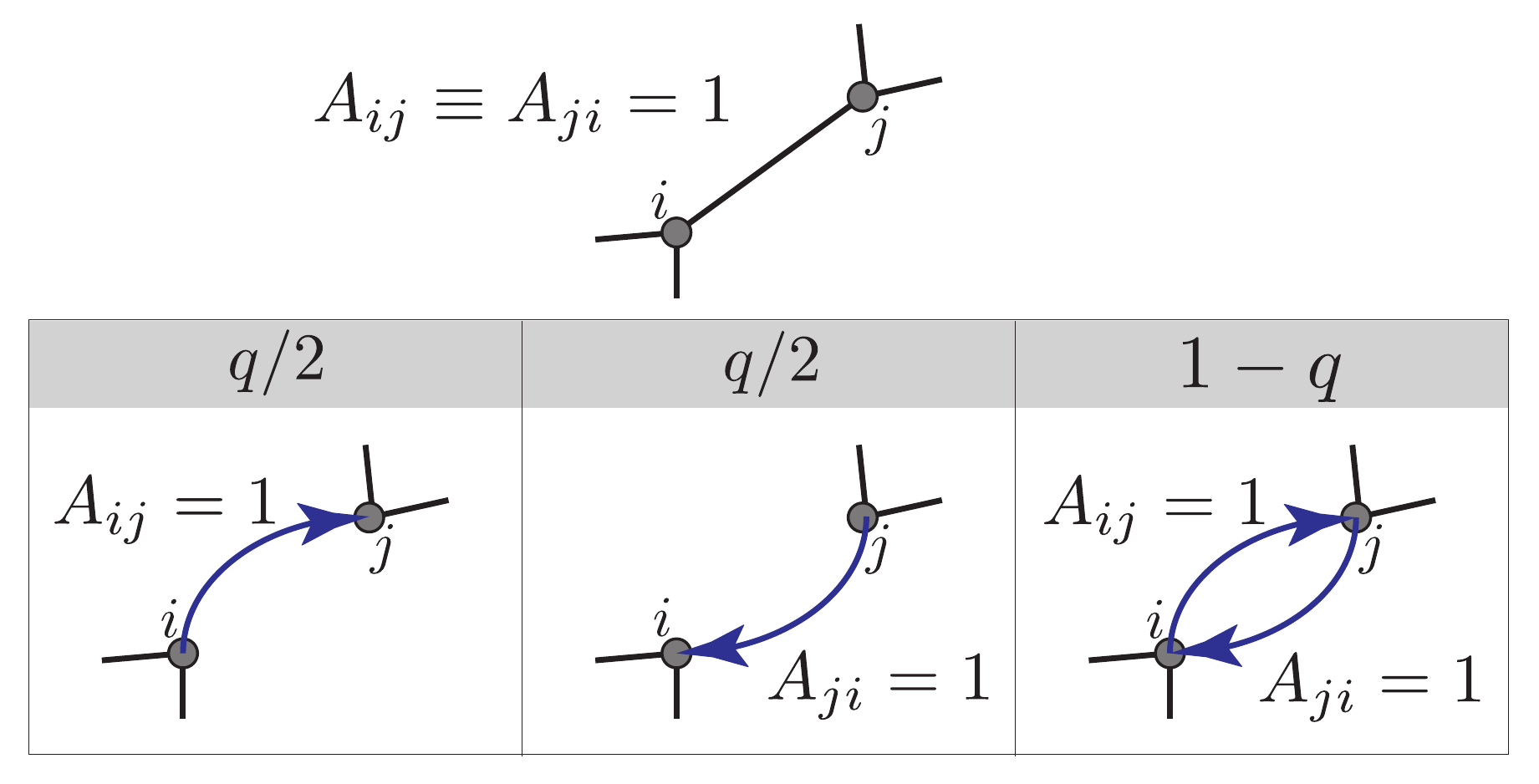}
    \caption{Schematic picture of uniformly random directization.}
    \label{fig-illust-random-directization}
\end{figure}

An undirected graph has a symmetric adjacency matrix $\bm{A}$, in which $A_{ij}=A_{ji}=1$ when vertices $i$ and $j$ are connected, while $A_{ij}=0$ otherwise.
The degree, defined by $d_i=\sum_j A_{ij} = \sum_j A_{ji}$, represents the number of the neighbors of vertex $i$.
For a directed graph, in contrast, the adjacency matrix is asymmetric, i.e., $A_{ij}=1$ and $A_{ji}=0$ when an edge has a direction from vertex $i$ to vertex $j$ only.
Hereafter, we regard an undirected edge in a directed graph as a pair of directed edges in both directions.
The in-degree and out-degree, defined by $d_i^{(\mathrm{in})}=\sum_j A_{ji}$ and $d_i^{(\mathrm{out})}=\sum_j A_{ij}$, denote the number of in-neighboring vertices (with in-coming edges) and out-neighboring vertices (with out-going edges) of vertex $i$, respectively.

We define uniformly random directization as follows.
Let $G(V,E)$ denote an undirected graph that consists of a set of vertices $V$ and edges $E$ and let $\tilde{G}(V,\tilde{E})$ denote a graph randomly directized from the original graph $G(V,E)$.
The number of vertices is denoted by $N=|V|$ for both graphs, while the numbers of edges for $G(V,E)$ and $\tilde{G}(V,\tilde{E})$ are respectively denoted by $M=|E|$ and $\tilde{M}=|\tilde{E}|$; note that each undirected edge is doubly counted in the latter, but not in the former.
Hereafter, $\bm{A}^{0}$ and $\tilde{\bm{A}}$ denote the adjacency matrices of $G(V,E)$ and $\tilde{G}(V,\tilde{E})$, respectively.
On directization, we alter an undirected edge $e_{ij}\in E$ between vertex $i$ and vertex $j$ to a directed edge $e_{i\rightarrow j}\in \tilde{E}$ (from $i$ to $j$) with probability $q/2$, to $e_{i\leftarrow j}\in \tilde{E}$ (from $j$ to $i$) with probability $q/2$, and remain undirected with probability $1-q$, where $0\leq q\leq 1$; see Fig.~\ref{fig-illust-random-directization}.
The number of directed edges after the uniformly random directization becomes $\tilde{M}=qM+2(1-q)M$ on average because we doubly count an undirected edge as a pair of directed edges with two directions.
We express the adjacency matrix $\tilde{\bm{A}}$ of the uniformly directized graph as 
\begin{equation}
\tilde{\bm{A}}=\bm{A}^{0}-\bm{V},
\label{eq-sec2-1}
\end{equation}
where $\bm{V}$ is the perturbation matrix defined as follows: when $A^0_{ij}\equiv A^0_{ji}=1$, the set of matrix elements $\left\{V_{ij}, V_{ji}\right\}$ takes $\left\{0, 1\right\}$, $\left\{1, 0\right\}$, or $\left\{0, 0\right\}$ with probabilities $q/2$, $q/2$, and $1-q$, respectively; otherwise, $V_{ij}=V_{ji}=0$.
We express the $i$th eigenvalue of $\bm{A^0}$ as $\lambda_i$ and its eigenvector as $\bm{v}_i$. 
The corresponding eigenvalue and eigenvector for $\tilde{\bm{A}}$ are denoted by $\tilde{\lambda}_i$ and $\tilde{\bm{v}}_i$, respectively.
We express the ensemble of perturbation matrices $\bm{V}$ for a given adjacency matrix $\bm{A}^0$ as $\mathcal{S}(\bm{A}^0)$.
Throughout this paper, the norm of each eigenvector is normalized to unity. 

\subsection{\label{section-2-1} Perturbative analysis}

Perturbation theory allows us to estimate the variation of each eigenvalue $\delta\lambda_i=\tilde{\lambda}_i-\lambda_i$ and the variation of each eigenvector $\delta \bm{v}_i=\tilde{\bm{v}}_i-\bm{v}_i$ under perturbative directization. 
We calculate the ensemble average and variance of these variations up to the first-order approximation.

\subsubsection{\label{section-2-1-1} Eigenvalues}

For a given graph, the variation of the $i$th eigenvalue $\lambda_i$ along the real axis in the first-order approximation \cite{sakurai2017modern,kato2013perturbation} is
\begin{equation}
\delta \lambda_i^{(1)}= -\bm{v}_i^{\top} \bm{V} \bm{v}_i,
\end{equation}
where $\bm{v}^{\top}$ denotes the transpose of a vector $\bm{v}$.
Note that this is for a specific instance of graph, whereas we are interested in the ensemble of randomly directized graphs. To this end, 
we define a generating function for $\delta \lambda_i^{(1)}$ by
\begin{align}
Z_i^{(1)}(\beta)
&=\notag 
\left[ e^{-\beta \bm{v}_i^{\top} \bm{V} \bm{v}_i} \right]_{\bm{V}|\bm{A}^0}\\
&=\label{eq3-eigval-genfunc} 
\left[\prod_{\ell<m} e^{- \beta v_{i\ell}v_{im}\left(V_{\ell m}+V_{m\ell}\right) } \right]_{\bm{V}|\bm{A}^0},
\end{align}
where $\beta$ is an auxiliary parameter, $v_{i\ell}$ denotes the $\ell$th element of $\bm{v}_i$, and 
$[\cdots]_{\bm{V}|\bm{A}^0}$ represents the random average over the ensemble $\mathcal{S}(\bm{A}^0)$ defined in
\begin{equation}
\left[f\left(\bm{V}\right)\right]_{\bm{V}|\bm{A}^0}=
\frac{1}{|\mathcal{S}(\bm{A}^0)|}
\sum_{\bm{V}\in\mathcal{S}(\bm{A}^0)} \mathrm{Prob}\left(\bm{V}\right) f\left(\bm{V}\right).
\end{equation}
We find the average and variance of the variation for $\lambda_i$ respectively by
\begin{align}
\left[\delta \lambda_i^{(1)}\right]_{\bm{V}|\bm{A}^0}
&=
\frac{\partial}{\partial \beta}\ln Z_i^{(1)}|_{\beta=0},\\
\sigma^2\left(\delta \lambda_i^{(1)}\right)_{\bm{V}|\bm{A}^0}
&=
\frac{\partial^2}{\partial \beta^2}\ln Z_i^{(1)}|_{\beta=0}.
\end{align}

Because each edge is directized independently, the random average in Eq.~(\ref{eq3-eigval-genfunc}) is calculated as
\begin{equation}
Z_i^{(1)}=
\mathop{\prod_{\ell<m}}_{\left(A^0_{\ell m}=1\right)} 
\left( 1-q +q e^{-\beta v_{i\ell}v_{im}}\right).
\end{equation}
Then, the average and variance of the variation $\delta \lambda_i$ are given by
\begin{align}
\left[\delta \lambda_i^{(1)}\right]_{\bm{V}|\bm{A}^0} 
&\label{eq-sec3-eigval-avr}=
-\frac{q}{2}\sum_{\ell,m} A^0_{\ell m} v_{i\ell}v_{im} 
 = -\frac{q\lambda_i}{2},\\
\sigma^2(\delta \lambda_i^{(1)})_{\bm{V}|\bm{A}^0}
&\label{eq-sec3-eigval-var}=
\frac{q\left(1-q \right)}{2}
\sum_{\ell,m} A^0_{\ell m}
v_{i\ell}^2 v_{im}^2,
\end{align}
respectively, where we used the fact
\begin{equation}
\sum_{\ell,m} A^0_{\ell m} v_{i\ell}v_{im}
= \bm{v}_i^{\top} \bm{A}^0 \bm{v}_i =\lambda_i.
\end{equation}

Equation~(\ref{eq-sec3-eigval-avr}) indicates that, in the range where the first-order approximation is valid, the average of the perturbed eigenvalue is 
\begin{equation}
\left[\tilde{\lambda}_i\right]_{\bm{V}|\bm{A}^0} = \lambda_i+\left[\delta \lambda_i^{(1)}\right]_{\bm{V}|\bm{A}^0}=\left(1-\frac{q}{2}\right) \lambda_i.
\end{equation}
Thus, the relative spectral structure is conserved under uniformly random directization as long as the first-order approximation is valid.

Let us investigate the condition that the fluctuation of the variation is small.
We consider the ratio between the average and standard deviation for the $i$th eigenvalue:
\begin{equation}
\frac{\sqrt{\sigma^2\left(\delta\lambda_i^{(1)}\right)_{\bm{V}|\bm{A}^0}}}
{\left|\left[\delta\lambda_i^{(1)}\right]_{\bm{V}|\bm{A}^0}\right|}=
\frac{1}{|\lambda_i|}
\sqrt{\frac{2}{q\left(1-q\right) } \sum_{\ell,m} A_{\ell m}^0 v_{i\ell}^2 v_{i m}^2}.
\end{equation}
Because each element of the eigenvector typically scales as $O\left(N^{-1/2}\right)$ and the number of nonzero elements in the summation is $cN$, where $c$ denotes the average degree of the original undirected graph, defined by $c=\sum_i d_i/N$, the ratio above scales as
\begin{equation}
\frac{\sqrt{\sigma^2\left(\delta\lambda_i^{(1)}\right)_{\bm{V}|\bm{A}^0}}}
{\left|\left[\delta\lambda_i^{(1)}\right]_{\bm{V}|\bm{A}^0}\right|}=
O\left(\frac{1}{|\lambda_i|} \sqrt{\frac{2c}{q \left(1-q\right) N}}\right).
\end{equation}
In the case of regular random graphs, the eigenvalue at the edge of the spectral band is $2\sqrt{c}$ \cite{mckay1981expected}.
Thus, the fluctuation of the variation is expected to be negligible when $2q \left(1-q\right) N$ is sufficiently large for the eigenvalues out of the spectral band.

\begin{figure*}[t!]
    \begin{tabular}{cc}
    \begin{minipage}[b]{.45\linewidth}
    \centering
    \includegraphics[width=\linewidth]{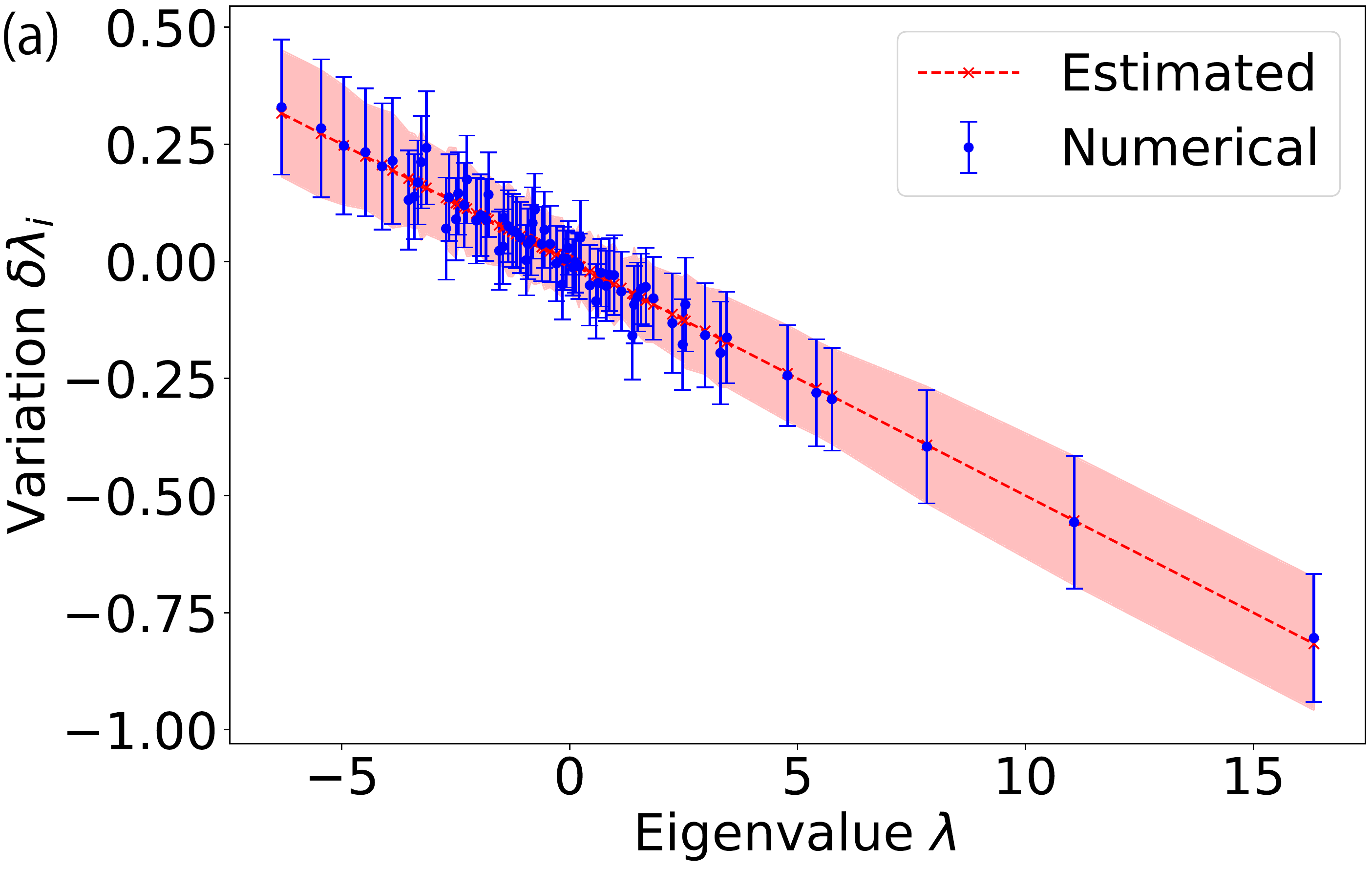}
    \end{minipage}
    &
    \begin{minipage}[b]{.45\linewidth}
    \centering
    \includegraphics[width=\linewidth]{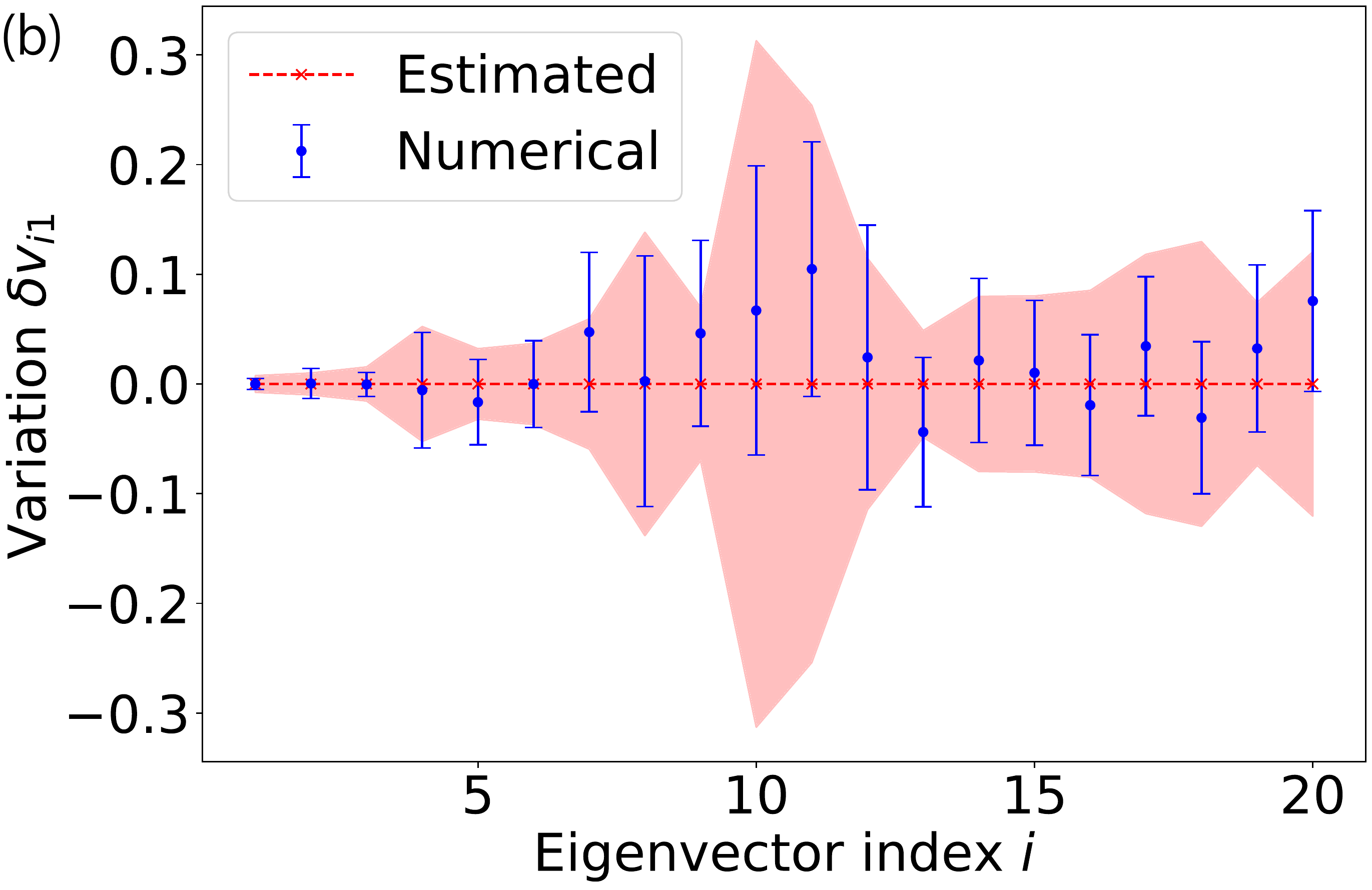}
    \end{minipage}
    \end{tabular}
    \caption{Variation of the eigenvalues and eigenvector elements of the adjacency matrix for the undirectized macaque-cortex network \cite{young1993organization} under uniformly random directization with $q=0.1$. Blue points and error bars indicate the average and standard deviation over 10,000 directized samples, respectively. (a) Variation of the real part of the eigenvalues. The horizontal axis indicates the eigenvalues of the unperturbed matrix $\bm{A}^0$.  The red dashed line and belt represent the estimates obtained using Eqs.~(\ref{eq-sec3-eigval-avr}) and (\ref{eq-sec3-eigval-var}), respectively. (b) Variation of the real part of the first element of each eigenvector $v_{i1}$. The red dashed line and belt represent the estimates obtained using Eqs.~(\ref{eq-sec3-eigvec-avr}) and (\ref{eq-sec3-eigvec-var}), respectively.}
    \label{fig-sec3-1}
\end{figure*}

\begin{figure*}[t!]
    \begin{tabular}{cc}
    \begin{minipage}[b]{.45\linewidth}
    \centering
    \includegraphics[width=\linewidth]{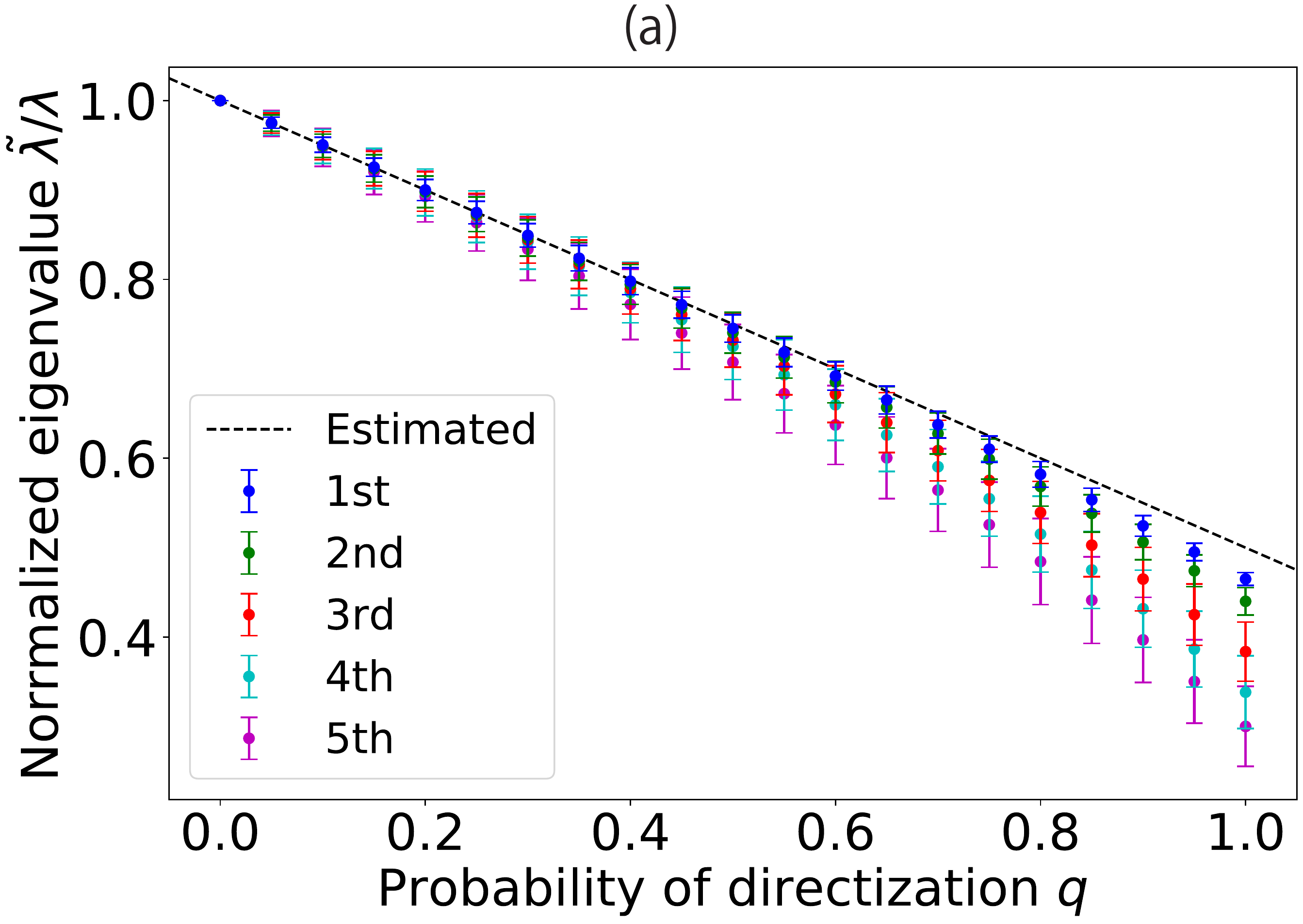}
    \end{minipage}
    &
    \begin{minipage}[b]{.45\linewidth}
    \centering
    \includegraphics[width=\linewidth]{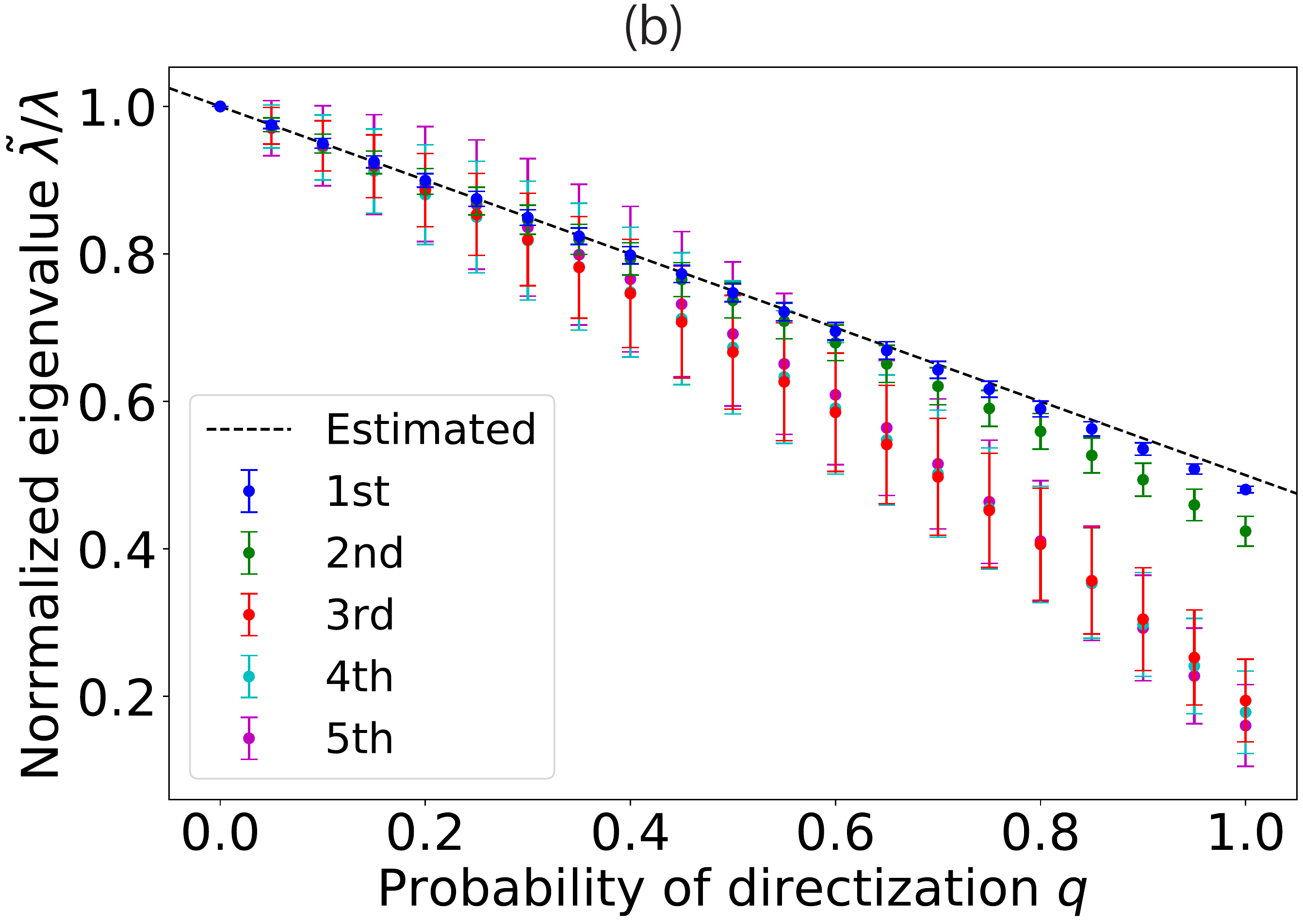}
    \end{minipage}
    \end{tabular}
    \caption{Normalized eigenvalues along the real axis of two real-world networks after uniformly random directization. Each point and error bar indicate the average and standard deviation over 10,000 directized samples, respectively. The dashed lines indicate the theoretical estimates obtained from Eq.~(\ref{eq-sec3-eigval-avr}). (a) Macaque-cortex network \cite{young1993organization}. (b) Social network of employees at a consulting company \cite{cross2004hidden}.}
    \label{fig-sec3-3}
\end{figure*}

\subsubsection{\label{section-2-1-2} Eigenvectors}

For a given graph, the variation of the $i$th eigenvector $\bm{v}_i$ along the real axis in the first-order approximation \cite{sakurai2017modern,kato2013perturbation} is
\begin{equation}
\delta\bm{v}_i^{(1)}
=-\sum_{j\neq i} 
\frac{\bm{v}^{\top}_j \bm{V} \bm{v}_i}{\lambda_i -\lambda_j} \bm{v}_j.
\end{equation}
For the $\ell$th element of $\bm{v}_i$, we define its generating function by
\begin{equation}
Z_{i\ell}^{(1)}=
\left[ 
e^{ -\beta \sum_{j\neq i}\frac{\bm{v}_j^{\top} \bm{V}\bm{v}_i}{\lambda_i-\lambda_j}v_{j\ell} }
\right]_{\bm{V}|\bm{A}^0}.
\end{equation}
We find the average and variance of the variation of the $\ell$th element of $\bm{v}_i$ respectively by
\begin{align}
\left[\delta v_{i\ell}^{(1)}\right]_{\bm{V}|\bm{A}^0}
&=
\frac{\partial}{\partial \beta} \ln Z_{i\ell}^{(1)}|_{\beta=0}, \label{eq-sec3-eigvec-avr0}\\
\sigma^2\left(\delta v_{i\ell}^{(1)}\right)_{\bm{V}|\bm{A}^0}
&=
\frac{\partial^2}{\partial \beta^2} \ln Z_{i\ell}^{(1)}|_{\beta=0}. \label{eq-sec3-eigvec-var0}
\end{align}
As was done for the eigenvalues above, we can take the random average with respect to each edge, obtaining
\begin{align}
Z_{i\ell}^{(1)} &=\notag
\prod_{j\neq i} 
\mathop{\prod_{m<n}}_{\left(A^0_{mn}=1\right)}  
 \Biggl(
 1-q  \\ &\quad \label{eq-sec3-eigvec-z}
+\frac{q}{2} e^{ - \frac{\beta v_{j\ell} v_{jm}v_{in}}{\lambda_i-\lambda_j} }
+\frac{q}{2} e^{ - \frac{\beta v_{j\ell} v_{jn}v_{im}}{\lambda_i-\lambda_j} }
\Biggr). 
\end{align}
From Eqs.~(\ref{eq-sec3-eigvec-avr0}) and (\ref{eq-sec3-eigvec-z}), we find the average of $\bm{v}_i$'s variation as follows:
\begin{align}
\left[\delta v_{i\ell}^{(1)}\right]_{\bm{V}|\bm{A}^0}
&=\notag
- \frac{q}{2} \sum_{j\neq i} \sum_{m,n} A^0_{mn}
\frac{v_{j\ell} v_{jm} v_{in}}{\lambda_i -\lambda_j} \\
&=\notag
- \frac{q}{2} \sum_{j\neq i} 
\frac{v_{j\ell}}{\lambda_i -\lambda_j} \bm{v}_j^{\top} \bm{A}^0 \bm{v}_i\\
&=\label{eq-sec3-eigvec-avr}
0.
\end{align}
Thus, in the first-order perturbative regime, uniformly random directization does not vary the eigenvectors of adjacency matrices on average.
The variance of $\bm{v}_i$'s variation is also obtained by using Eqs.~(\ref{eq-sec3-eigvec-var0}) and (\ref{eq-sec3-eigvec-z}): 
\begin{align}
\sigma^2\left(\delta v_{i\ell}^{(1)} \right)_{\bm{V}|\bm{A}^0}
&=\notag
\sum_{j\neq i} 
\sum_{m,n} A^0_{mn}  
\Biggl(
\frac{q}{2}
\left(\frac{v_{j\ell} v_{jm} v_{in}}{\lambda_i -\lambda_j}\right)^2\\
&-
\left(\frac{q}{2}\right)^2 \frac{1}{2} 
\left(\frac{v_{j\ell} v_{jm} v_{in} + v_{j\ell} v_{jn} v_{im}}{\lambda_i -\lambda_j}\right)^2
\Biggr)
\label{eq-sec3-eigvec-var}
\end{align}

\subsection{\label{section-2-2} Numerical confirmation}

We now compare the first-order perturbative estimation with numerical calculations.
We generated 10,000 uniformly randomly directized samples from the undirectized counterpart for real-world networks, and numerically calculated the eigenvalues and eigenvectors of the adjacency matrices.

Figure~\ref{fig-sec3-1} exhibits the variation of the real part of each eigenvalue $\delta\lambda_{i}$ and the first element of each eigenvector $\delta v_{i1}$ of the undirectized macaque-cortex network \cite{young1993organization} under random directization.
We show the variation of all eigenvalues in the left panel and the variation of eigenvector elements corresponding to the top 20 eigenvalues in the right panel.
In both panels, we observe that the first-order perturbative estimates and the numerical results agree well, particularly for the top eigenvalues, which are expected to be isolated eigenvalues.

Figure~\ref{fig-sec3-3} shows the $q$-dependency of the top five eigenvalues along the real axis for two real-world directed networks; the macaque cortex network \cite{young1993organization} and social network of employees at a consulting company \cite{cross2004hidden}.
The fractions of directed edges are 0.30 and 0.41, respectively.
We numerically calculated the perturbed eigenvalues normalized by the original eigenvalues.
Based on Eq.~(\ref{eq-sec3-eigval-avr}), we estimate the normalized eigenvalue $\tilde{\lambda}_i/\lambda_i$ to be $1-q/2$ regardless of the index of eigenvalue, which is illustrated by the dashed line in Fig.~\ref{fig-sec3-3}.
For both networks, the estimation is reasonable when $q$ is sufficiently small.
As $q$ increases, the difference between the theoretical estimate and the numerical result increases.

\begin{figure}[t!]
    \includegraphics[width=\linewidth]{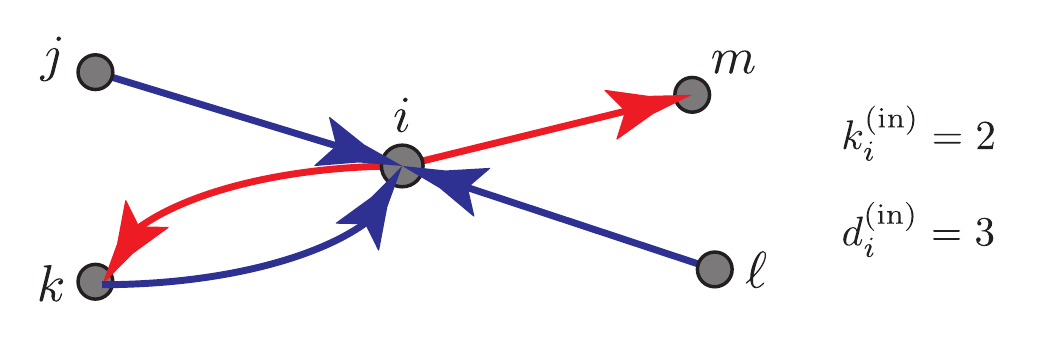}
    \caption{In this example, the directed in-degree is $k_i^{\mathrm{(in)}}=2$, for which we count $e_{j \to i}$ and $e_{\ell \to i}$, whereas the in-degree is $d_i^{\mathrm{(in)}}=3$, for which we count $e_{j \to i}$, $e_{\ell \to i}$, and $e_{k \to i}$.}
    \label{fig:illust-directed-indegree}
\end{figure}

\begin{figure}[t!]
    \includegraphics[width=\linewidth]{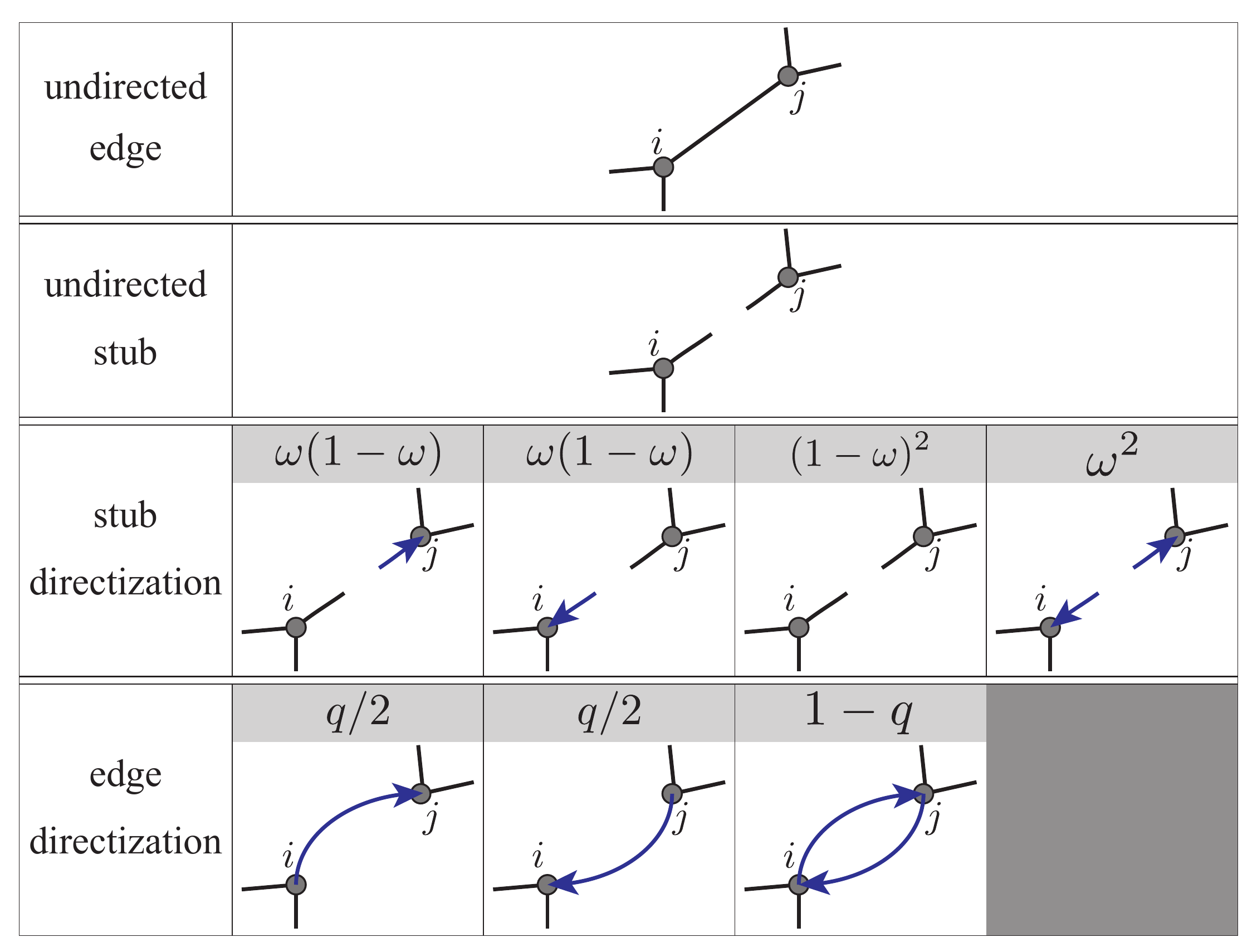}
    \caption{Schematic picture of random directization of stubs.}
    \label{fig:illust-stub-directization}
\end{figure}

\begin{figure*}[t!]
    \begin{tabular}{cc}
    \begin{minipage}[b]{.70\linewidth}
    \centering
    \includegraphics[width=\linewidth]{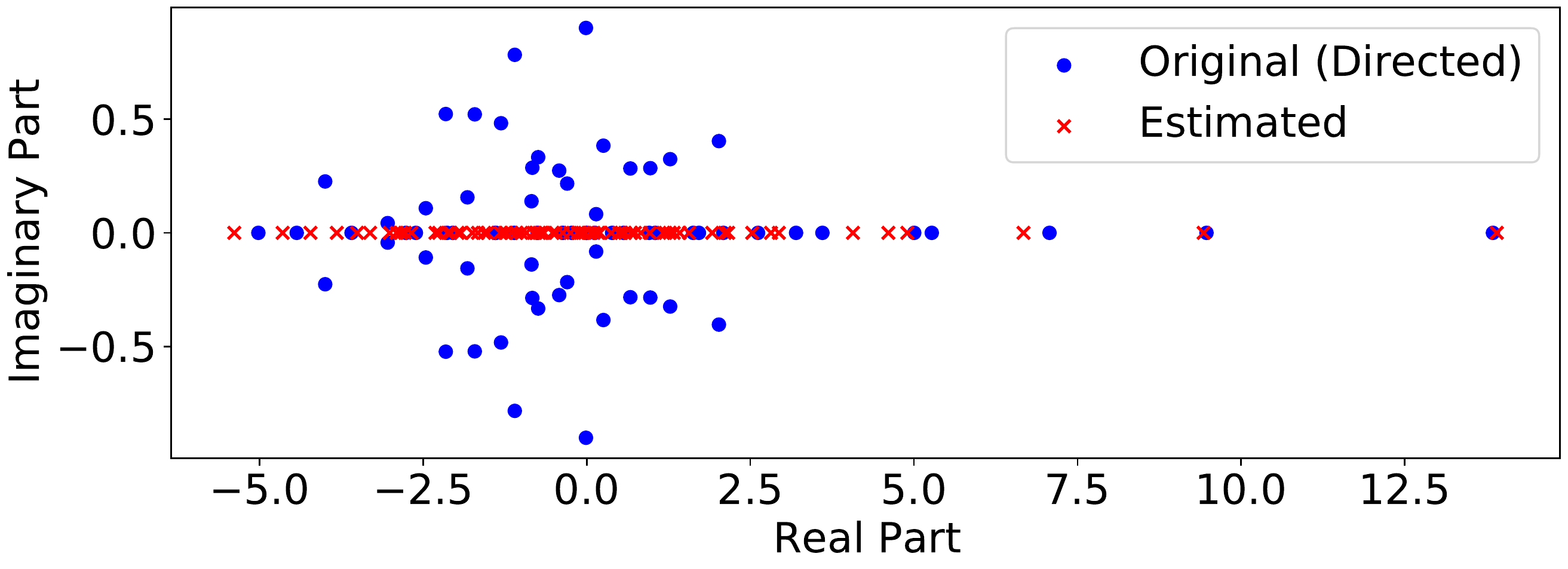}
    \end{minipage}
    &
    \begin{minipage}[b]{.25\linewidth}
    \centering
    \includegraphics[width=\linewidth]{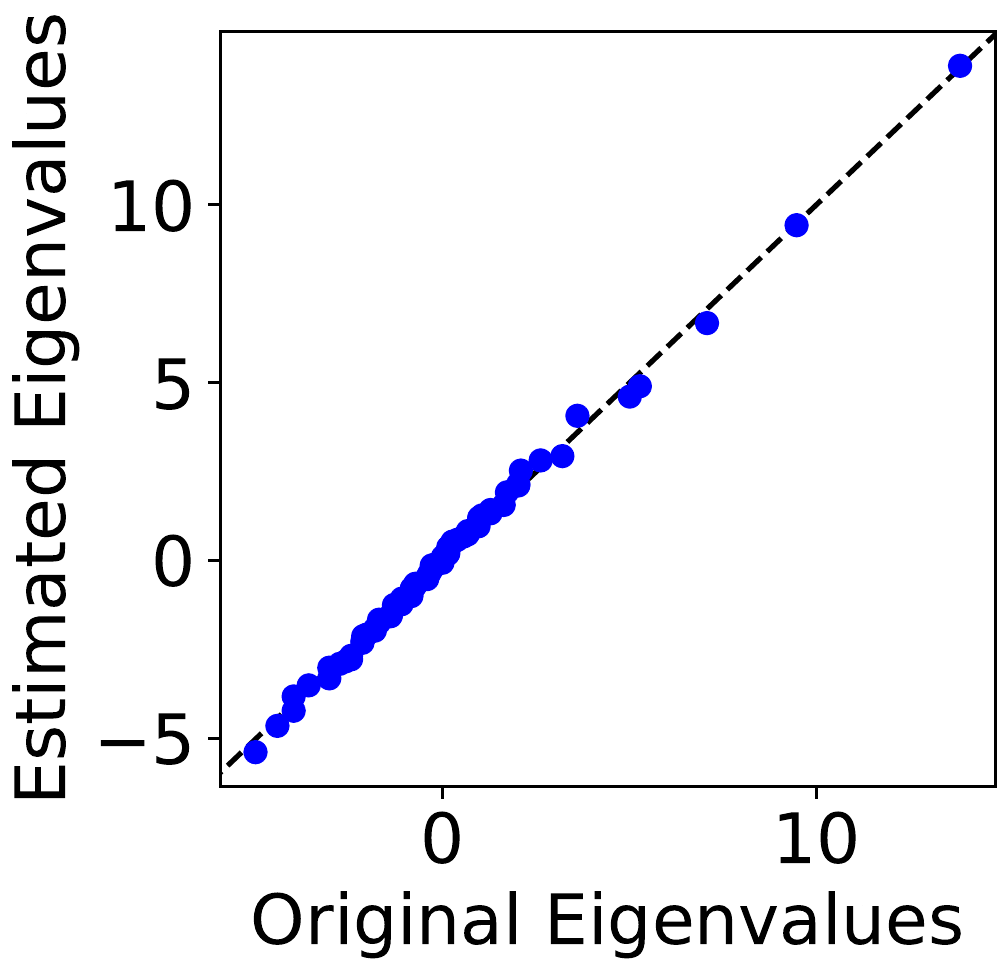}
    \end{minipage}\\
    \begin{minipage}[b]{.70\linewidth}
    \centering
    \includegraphics[width=\linewidth]{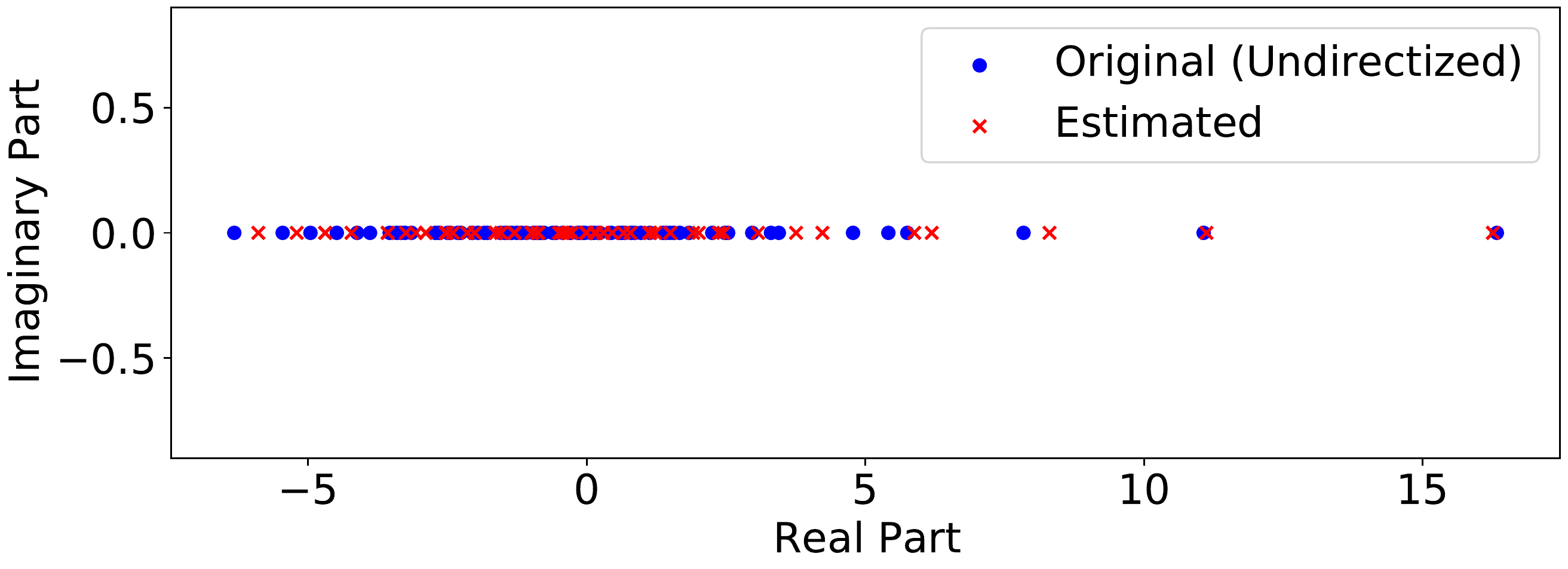}
    \end{minipage}
    &
    \begin{minipage}[b]{.25\linewidth}
    \centering
    \includegraphics[width=\linewidth]{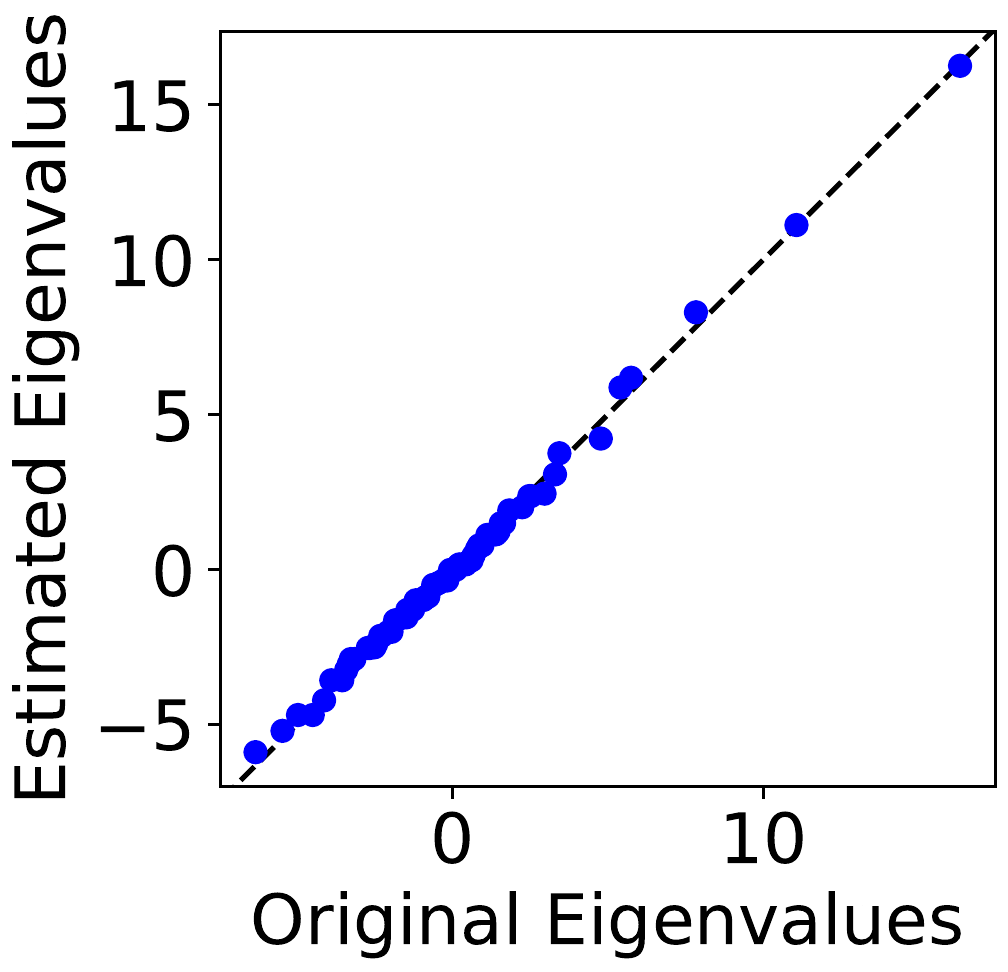}
    \end{minipage}
    \end{tabular}
    \caption{Adjacency matrix spectra of the macaque-cortex network. 
    In the top panels, we compare the spectrum of the original directed graph (blue points) with the result of uniformly random directization from Eq.~(\ref{eq-sec3-eigval-avr}) (red crosses). 
    In the bottom panels, we compare the spectrum of the undirectized graph (blue points) with the estimates from Eq.~(\ref{eq-sec-4}) (red crosses).  
    In each of the right panels, the real parts of the original and estimated eigenvalues are compared.
    The points are located near the dashed diagonal line, indicating that these values coincide well.}
    \label{fig-sec4}
\end{figure*}

\subsection{\label{section-2-3} Distribution of directed in-degrees}

Before we conclude this section, here we consider the degree distribution of directized edges after uniformly random directization.
We define the \textit{directed in-degree} $k_i^{(\mathrm{in})}$ for a directed graph as the number of in-coming edges for each vertex:
\begin{equation}
k_i^{(\mathrm{in})} = \mathop{\sum_{j=1}^N}_{\left(A_{ij}=0\right)} A_{ji} 
= d_i - d_i^{(\mathrm{out})} = \sum_{j=1}^N V_{ij}
\end{equation}
Here, we do not count undirected edges for $k_i^{(\mathrm{in})}$, whereas we count them for $d_i^{(\mathrm{in})}$.
A simple example is shown in Fig.~\ref{fig:illust-directed-indegree}.

Note that because the directization of an edge affects the directed in-degrees of the vertices on both ends, the directed in-degrees of neighboring vertices are correlated with each other. 
Instead of deriving the directed in-degree distribution after actual random directization, we derive it after a process in which we uniformly randomly directize stubs (half-edges).
This procedure is illustrated in Fig.~\ref{fig:illust-stub-directization}.
We independently alter an undirected stub to the in-coming one with probability $\omega$ and keep it undirected with probability $1-\omega$.
The probability that the number of directed stubs for vertex $i$ is $k$ is given by a binomial distribution:
\begin{equation}
p^{\textrm{stub}}_i(k)=\binom{d_i}{k} \omega^k \left(1-\omega \right)^{d_i-k}.
\end{equation}
We regard that each edge becomes a directed edge if one end of the edge is directized while the other end is not.
After random directization of stubs, an edge $e_{ij}$ is converted to $e_{i\rightarrow j}$ with probability $\omega (1-\omega)$, converted to $e_{i\leftarrow j}$ with probability $\omega (1-\omega)$,  and remains undirected with probability $(1-\omega)^2$.
With probability $\omega^{2}$, an edge becomes bi-directed and the resulting object is no longer a directed graph that we consider.
Nonetheless, when $\omega$ is sufficiently small such that the emergence of the bi-directed edges is negligibly rare, this process is almost identical to the directization defined in Eq.~(\ref{eq-sec2-1}) with $\omega = q/2$.
Thus, when $q$ is sufficiently small, the probability that a vertex $i$ has a directed in-degree $k_i^{(\mathrm{in})}$ after uniformly random directization approximately follows a binomial distribution:
\begin{equation}
p_i(k_i^{(\mathrm{in})})\simeq
\binom{d_i}{k_i^{(\mathrm{in})}} 
\left(\frac{q}{2}\right)^{k_i^{(\mathrm{in})}} \left(1-\frac{q}{2} \right)^{d_i-k_i^{(\mathrm{in})}}.
\end{equation}

Then, we obtain the directed in-degree distribution over a graph after uniformly random directization.
For a subset of vertices with the same degree $d$, the directed in-degree distribution approximately follows a binomial distribution:
\begin{equation}
P_d(k^{(\mathrm{in})})\simeq\binom{d}{k^{(\mathrm{in})}} \left(\frac{q}{2}\right)^{k^{(\mathrm{in})}} \left(1-\frac{q}{2} \right)^{d-k^{(\mathrm{in})}}.
\end{equation}
Thus, the directed in-degree distribution $P(k^{(\mathrm{in})})$ over the whole graph approximately follows the mixture of binomial distributions:
\begin{align}
P(k^{(\mathrm{in})})&=\notag \sum_{d=0}^{\infty} Q(d) P_d(k^{(\mathrm{in})})\\
&\simeq \sum_{d=0}^{\infty} Q(d) \binom{d}{k^{(\mathrm{in})}} \left(\frac{q}{2}\right)^{k^{(\mathrm{in})}} \left(1-\frac{q}{2} \right)^{d-k^{(\mathrm{in})}},
\label{eq-sec4-2}
\end{align}
where $Q(d)$ denotes the degree distribution of the original undirected graph.

\section{\label{section-3} Variation of spectra under undirectization}

We return to our original motivation of clarifying how the graph spectra are varied by ignoring edge directions, namely undirectization.
Let $G^{\mathrm{D}}(V,E^{\mathrm{D}})$ denote a directed graph and let $G^{\mathrm{U}}(V,E^{\mathrm{U}})$ denote its undirectized graph.
The numbers of edges for $G^{\mathrm{D}}$ and $G^{\mathrm{U}}$ are respectively denoted by $M^{\mathrm{D}}=|E^{\mathrm{D}}|$ and $M^{\mathrm{U}}=|E^{\mathrm{U}}|$.
The fraction of directed edges, $\hat{q}$, which corresponds to the probability of directization in random directization, should satisfy $M^{\textrm{D}}=\hat{q} M^{\textrm{U}} + 2(1-\hat{q}) M^{\textrm{U}}$, and hence is given by $\hat{q}=2-M^{\mathrm{D}}/M^{\mathrm{U}}$.
The adjacency matrices for $G^{\mathrm{D}}$ and $G^{\mathrm{U}}$ are respectively denoted by $\bm{A}^{\mathrm{D}}$ and $\bm{A}^{\mathrm{U}}$.

The result from random directization implies that the relative spectral structure along the real axis is typically maintained under undirectization when the fraction $\hat{q}$ is sufficiently small.
The result of directization, Eq.~(\ref{eq-sec3-eigval-avr}), conversely implies that, after ignoring the edge directions, the real part of the $i$th eigenvalue $\lambda_i^{\textrm{D}}$ of $\bm{A}^{\mathrm{D}}$ is altered to the corresponding eigenvalue $\lambda_i^{\textrm{U}}$ of $\bm{A}^{\mathrm{U}}$ as in
\begin{equation}
\lambda^{\mathrm{U}}_i\simeq \left(1-\frac{\hat{q}}{2}\right)^{-1} \mathrm{Re} \left[ \lambda^{\mathrm{D}}_i \right],
\label{eq-sec-4}
\end{equation}
where $\mathrm{Re} \left[\cdots\right]$ denotes the real part of a complex value.

The perturbative analysis explains why the two spectra shown in Fig.~\ref{fig-sec1-1} share almost the same relative spectral structure for the real parts.
In Fig.~\ref{fig-sec4}, we compare the spectra of the macaque-cortex network, in which $\hat{q}=0.30$, with the theoretical estimates.
The top panel shows the resulting spectrum after random directization in Eq.~(\ref{eq-sec3-eigval-avr}), and the bottom panel shows the resulting spectrum after undirectization in Eq.~(\ref{eq-sec-4}).
Both panels show that the perturbative analysis is moderately accurate, particularly for isolated eigenvalues.

\begin{figure}[t!]
    \includegraphics[width=\linewidth]{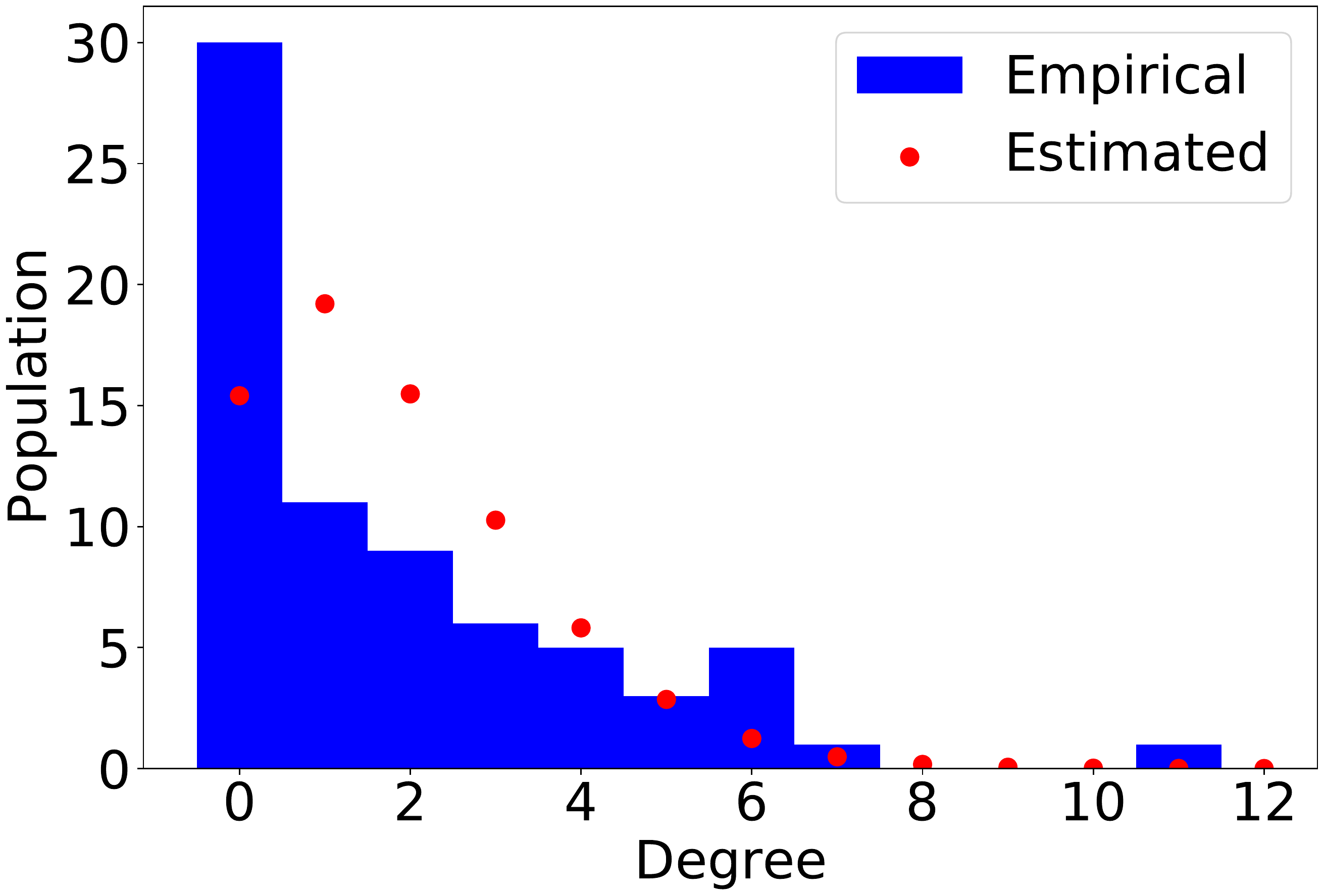}
    \caption{Empirical and estimated directed in-degree distributions of the macaque-cortex network. The red points represent the estimates from Eq.~(\ref{eq-sec4-2}).}
    \label{fig-sec4-2}
\end{figure}

Apart from the accuracy of the perturbation theory, we can assess, using Eq.~(\ref{eq-sec4-2}), to what degree a directed graph is
regarded as a uniformly randomly directized one.
Figure~\ref{fig-sec4-2} compares the directed in-degree distribution of the original (directed) macaque-cortex network and the theoretical prediction [Eq.~(\ref{eq-sec4-2})] for a uniformly randomly directed graph.
We assess the null hypothesis that the edges are directized uniformly randomly, by the $\chi^{2}$ goodness-of-fit test for the distributions in the range $k^{\mathrm{in}} \le 11$. 
As a result, we find that the $p$-value of the empirical distribution is less than $0.01$, which implies that the macaque-cortex network may not be regarded as a uniformly randomly directed graph, despite the accuracy of the perturbative analysis.

Note that the uniformity in random directization is not a necessary condition for the conservation of the relative spectral structure. 
Thus, the $\chi^{2}$ goodness-of-fit test of the in-degree distribution itself is not a criterion for the validity of our perturbative analysis. 
Nonetheless, this test partly explains why our analysis is plausible when the null hypothesis is not rejected.

\section{\label{section-4}Summary and discussions}

In this study, we investigated the contribution of directedness in graph spectra.
We introduced random directization as the inverse operation of ignoring the edge directions.
We revealed that, in the perturbative regime, uniformly random directization does not destroy the relative spectral structure along the real axis of the undirected graphs.
Additionally, we observed that the relative spectral structure along the real axis was also conserved for real-world datasets.
Although the effects of directization and undirectization on the graph spectra are generally not symmetric, we showed that the results of random directization can be used to explain the behavior of the undirectization.

\begin{figure*}[t!]
    \centering
    \includegraphics[width=\linewidth]{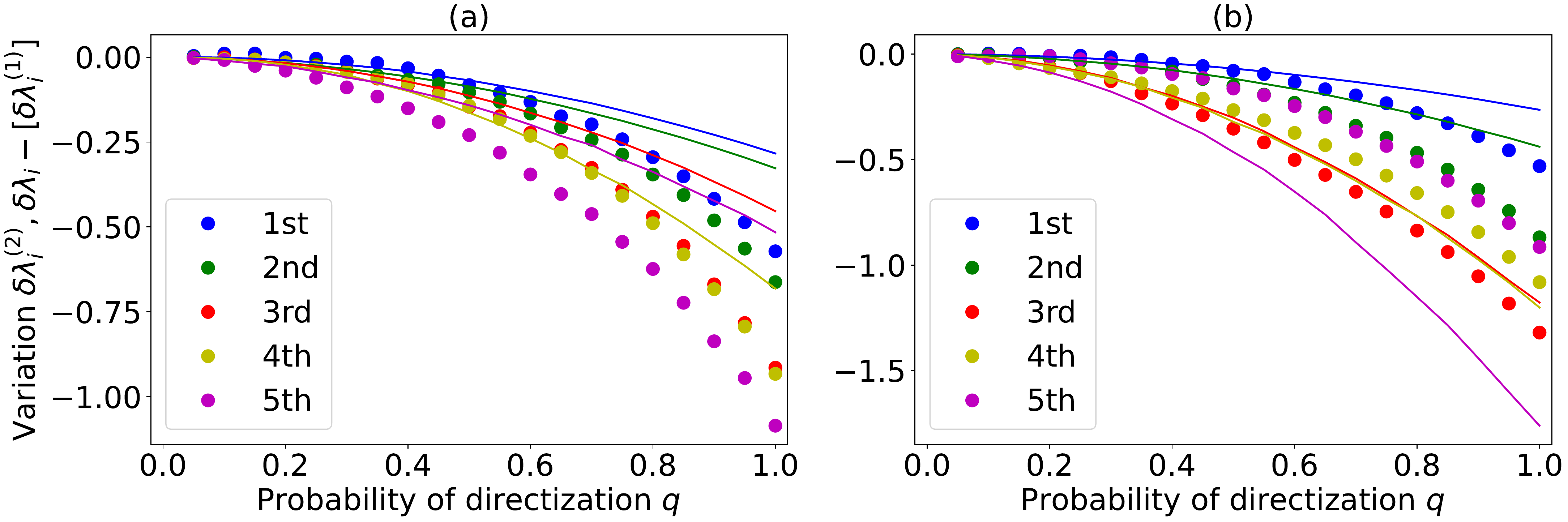}
    \caption{The contribution of higher order terms in the perturbative analysis. Each point indicates the real part of the deviation from the first-order estimates [Eq.~(\ref{eq-sec3-eigval-avr})] $\delta\lambda_i-\left[\delta\lambda_i^{(1)}\right]_{\bm{V}|\bm{A}^0}$. The solid lines indicate the random average of $\delta\lambda_i^{(2)}$ [Eq.~(\ref{eq-sec4-dl})] over the 10,000 samples. (a) Macaque-cortex network \cite{young1993organization}. (b) Social network of employees at a consulting company \cite{cross2004hidden}.}
    \label{fig-sec4-1}
\end{figure*}

Several comments are in order. 
In Sec.~\ref{section-2}, we analyzed up to the first-order term in the perturbative expansion.
However, it is not obvious whether the contributions from higher-order terms are negligible. 
In our formulation of perturbative random directization in Eq.~(\ref{eq-sec2-1}), $q$ is not a perturbative expansion parameter, but simply defines the fraction of non-zero elements.
Instead, we carry out the perturbative expansion with respect to the matrix $\bm{V}$, which does not consist of infinitesimally small elements. 
Here, let us consider the contribution from higher-order terms, especially the second-order term.
The second-order terms in the perturbative expansion of the $i$th eigenvalue $\lambda_i$ along the real axis \cite{sakurai2017modern,kato2013perturbation} is given by
\begin{align}
\delta\lambda_i^{(2)}=
\sum_{j\neq i} \frac{\bm{v}^{\top}_i \bm{V} \bm{v}_j \bm{v}^{\top}_j \bm{V} \bm{v}_i}{\lambda_i-\lambda_j}. \label{eq-sec4-dl}
\end{align}
We numerically calculate the average of this second-order contribution $\delta\lambda_i^{(2)}$ over randomly directized graphs and show the result in Fig.~\ref{fig-sec4-1} as solid lines.
We also show the real part of the deviation from the first-order approximation $\delta\lambda_i-\left[\delta\lambda_i^{(1)}\right]_{\bm{V}|\bm{A}^0}$ in Fig.~\ref{fig-sec4-1} as points.
We observe that the second-order contribution is smaller when $q$ is sufficiently small, and the first-order approximation is indeed valid in that region.
In addition, the average of the second-order term is not proportional to the original eigenvalues while that of the first-order term is.
Therefore, the spectral structure can be much more complicated when higher-order contributions are dominant.
In Appendix~\ref{sec:app1}, we analytically calculated the second-order perturbation expansion up to the first order in $q$. 
As an exceptional case, we can analytically obtain the random average of the variation for random graphs when $q=1$ using the cavity method.
We show the result for the stochastic block model in Appendix~\ref{sec:app2}.

Second, we can easily show that the conservation property of the relative spectral structure cannot be generalized to arbitrary directizations.
For example, let us consider non-uniform directization on a graph with a module structure with two blocks.
When a specific edge $e_{ij}$ is altered to be directed as $e_{i\rightarrow j}$, the variation of each eigenvalue is expressed as
\begin{equation}
\delta\lambda_{\ell}=-\bm{v}_{\ell}^{\top}\bm{V}\bm{v}_{\ell}=-v_{\ell i} v_{\ell j}.
\end{equation}
The largest eigenvalue $\lambda_1$ varies negatively, regardless of the choice of the edges $e_{ij}$ because $v_{1i}$ and $v_{1j}$ always have the same sign thanks to the Perron-Frobenius theorem.
On the other hand, the variation of the second-largest eigenvalue $\lambda_2$, which is related to the module structure, depends on which edge is altered.
When the edge $e_{ij}$ functions as a bridge between the two blocks, $\delta\lambda_2$ is expected to be positive because the eigenvector elements $v_{2i}$ and $v_{2j}$ typically have different signs.
In contrast, when both ends of the edge $e_{ij}$ are located inside a common block, $\delta\lambda_2$ is expected to be negative because $v_{2i}$ and $v_{2j}$ typically have the same sign.
Thus, the relative spectral structure is not conserved when the random directization is not uniformly random.

In this study, we evaluate the spectral variation only along the real axis.
In the perturbative regime, all eigengaps along the real axis remain finite; thus, complex conjugate pairs never appear.
As we further asymmetrize the adjacency matrix, some of the neighboring eigenvalues may collide  at an exceptional point \cite{kato2013perturbation} and turn into a pair of complex conjugate eigenvalues.
At the exceptional point, the corresponding eigenvectors become parallel to each other, and hence the matrix becomes undiagonalizable.
The perturbation theory would no longer be valid because the matrices are assumed to be diagonalizable.

Random directization presented in this paper is applicable to resampling of graphs.
Many complex systems have only one network data at one time, which makes it difficult to statistically analyze its properties, including the spectrum.
To address this problem, there have been several resampling methods to duplicate similar undirected graphs from the original graph \cite{rosvall2010mapping,bhattacharyya2015subsampling}.
The present study implies that we can utilize random directization for the purpose of resampling directed graphs with the same configuration of edges, whose relative spectral structure is typically close to the original undirected graphs.

\begin{acknowledgements}
The authors thank Naomichi Hatano for his fruitful comments.
This work was supported by JSPS KAKENHI No. 19H01506 (T.K. and M.O.), JST ACT-X Grant No. JPMJAX21A8 (T.K. and M.O.) and Quantum Science and Technology Fellowship Program (Q-STEP) (M.O.).
\end{acknowledgements}

\bibliography{Contribution_of_directedness_in_graph_spectra.bib}

\appendix

\newpage
\begin{widetext}

\section{\label{sec:app-notation} List of symbols}

Here, we show the list of notations used in the main article in Fig.~\ref{Fig:AppendixA}.

\begin{figure}[h!]
    \includegraphics[width=\linewidth]{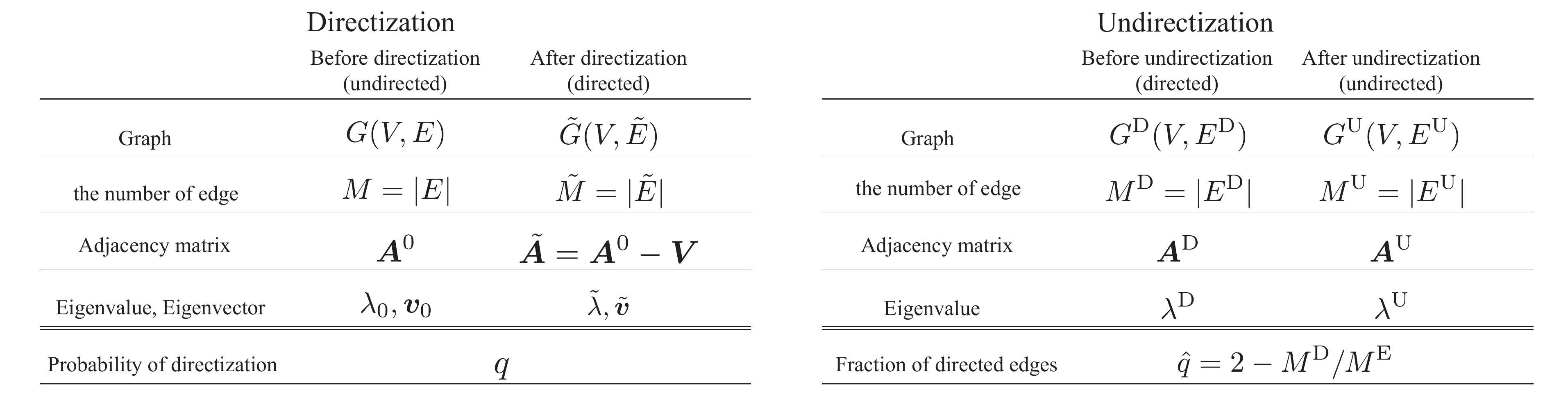}
    \caption{List of symbols used in this paper.}
    \label{Fig:AppendixA}
\end{figure}

\section{\label{sec:app1}Contribution of the second-order terms in the perturbation theory}

Here, we conduct an analytical calculation for the random average of the second-order terms of the eigenvalues and eigenvector elements up to the first order of $q$.
Note that, from the numerical calculation in Fig.~\ref{fig-sec4-1}, the random average of the second-order terms is not linear to $q$, and thus this form of first-order evaluation is only valid when $q$ is sufficiently small.

The second-order terms in the perturbation theory of the $i$th eigenvalue $\lambda_i$ and the corresponding eigenvector $\bm{v}_i$ along the real axis \cite{sakurai2017modern,kato2013perturbation} are, respectively, given by
\begin{align}
&\delta\lambda_i^{(2)}=\label{eq-app1-2ndeigval0}
\sum_{j\neq i} \frac{\bm{v}^{\top}_i \bm{V} \bm{v}_j \bm{v}^{\top}_j \bm{V} \bm{v}_i}{\lambda_i-\lambda_j}, \\
&\delta\bm{v}_i^{(2)}=\label{eq-app1-2ndeigvec0}
-
\frac{1}{2} \sum_{j\neq i} \left(\frac{\bm{v}_j^{\top} \bm{V} \bm{v}_i}{\lambda_i-\lambda_j}\right)^2 \bm{v}_i
+\sum_{j\neq i} \left( \sum_{k\neq i} 
\frac{\bm{v}^{\top}_j \bm{V} \bm{v}_k \bm{v}^{\top}_k \bm{V} \bm{v}_i}{\left(\lambda_i -\lambda_j\right) \left(\lambda_i -\lambda_k\right)}
-\frac{\bm{v}^{\top}_j \bm{V} \bm{v}_i \bm{v}^{\top}_i \bm{V} \bm{v}_i}{\left(\lambda_i -\lambda_j\right)^2}
\right) \bm{v}_j.
\end{align}
Similarly to the case of the first-order approximation, we respectively obtain the average of the second-order term up to the first order of $q$ for the $i$th eigenvalue and the $\ell$th element of the $i$th eigenvector $v_{i\ell}$ under uniformly random directization in the forms
\begin{align}
&\left[\delta\lambda_i^{(2)}\right]_{\bm{V}|\bm{A}^0}=
\frac{q}{2}\sum_{j \neq i} \sum_{m,n} A^0_{mn} \frac{v_{im} v_{in} v_{jm} v_{jn} }{\lambda_i-\lambda_j},
\label{eq-app1-2ndeigval}\\
&\left[\delta v_{i\ell}^{(2)}\right]_{\bm{V}|\bm{A}^0}=
-\frac{q}{2} \sum_{j\neq i}
\sum_{m,n} A_{mn}^0  
\left(
\frac{1}{2} \left(\frac{v_{jm}v_{in}}{\lambda_i-\lambda_j}\right)^2 v_{i\ell}
 +
\left(
-
\sum_{k\neq i} \frac{v_{jm} v_{kn} v_{km} v_{in}}{\left(\lambda_i -\lambda_j\right) \left(\lambda_i -\lambda_k\right)} 
+
\frac{v_{jm} v_{in} v_{im} v_{in}}{\left(\lambda_i -\lambda_j\right)^2}
\right) v_{j\ell}
\right).
\label{eq-app1-2ndeigvec}
\end{align}
The detailed calculations are shown in Appendix~\ref{sec:app4}.

We find that the second-order variation of an eigenvalue is not typically proportional to the original one as in Eq.~(\ref{eq-sec3-eigval-avr}) and that of the eigenvector elements do not vanish as in Eq.~(\ref{eq-sec3-eigvec-avr}).
Thus, the relative spectral structure is not conserved under uniformly random directization when the contribution of the second-order term is sufficiently large.

Figure~\ref{fig-app1-1} numerically compares the second-order terms, Eqs.~(\ref{eq-app1-2ndeigval}) and (\ref{eq-app1-2ndeigvec}), with the first-order terms, Eqs.~(\ref{eq-sec3-eigval-avr}) and (\ref{eq-sec3-eigvec-avr}), for the undirectized macaque-cortex network.
We observe that the second-order term is sufficiently smaller than the first-order term for some of the top eigenvalues.
On the other hand, in the region in which the eigenvalues gather densely around zero, the second-order term is comparable to the first-order term, and our first-order approximation is invalidated.

\begin{figure*}[t!]
    \begin{tabular}{cc}
    \begin{minipage}[b]{.45\linewidth}
    \centering
    \includegraphics[width=\linewidth]{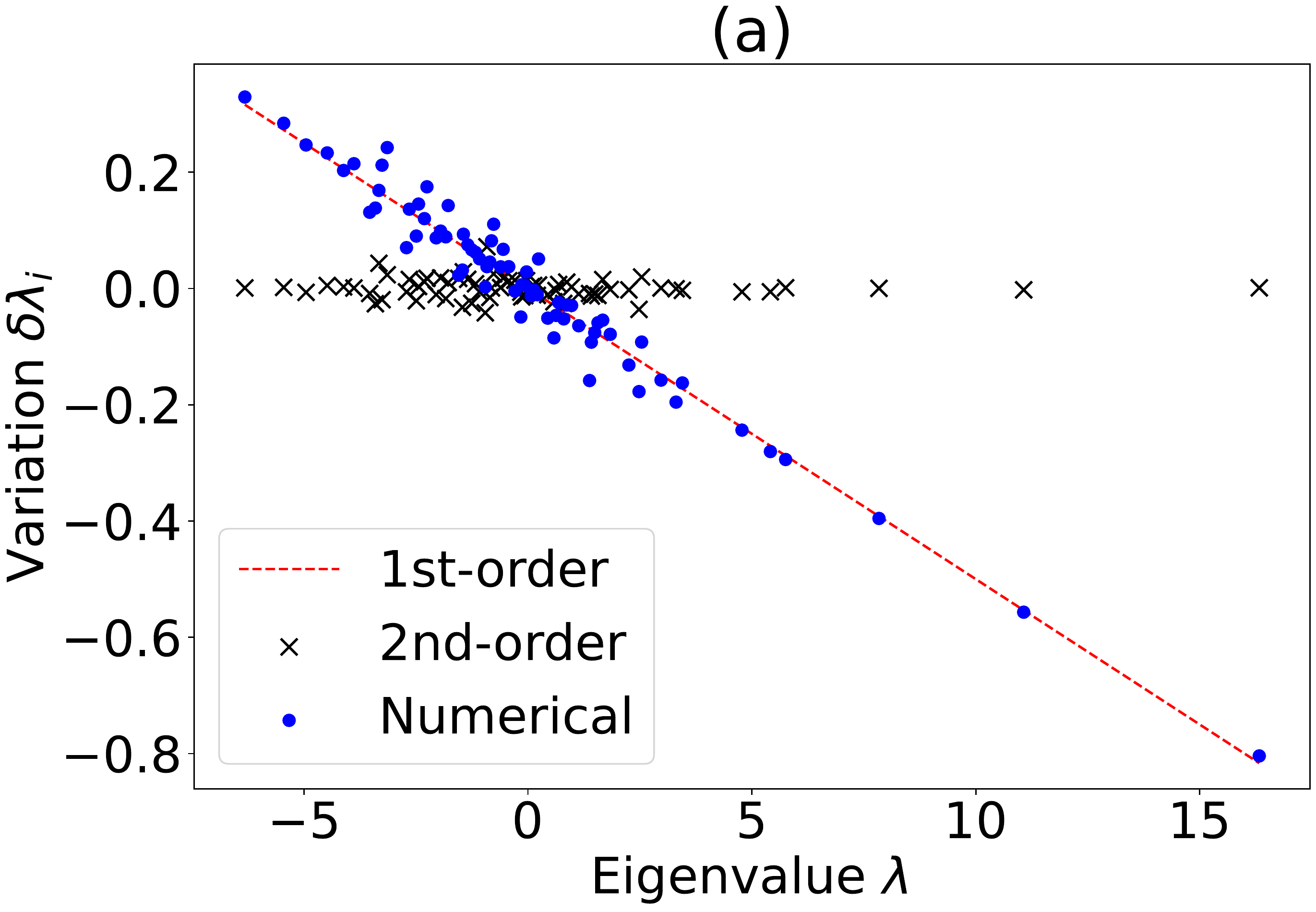}
    \end{minipage}
    &
    \begin{minipage}[b]{.45\linewidth}
    \centering
    \includegraphics[width=\linewidth]{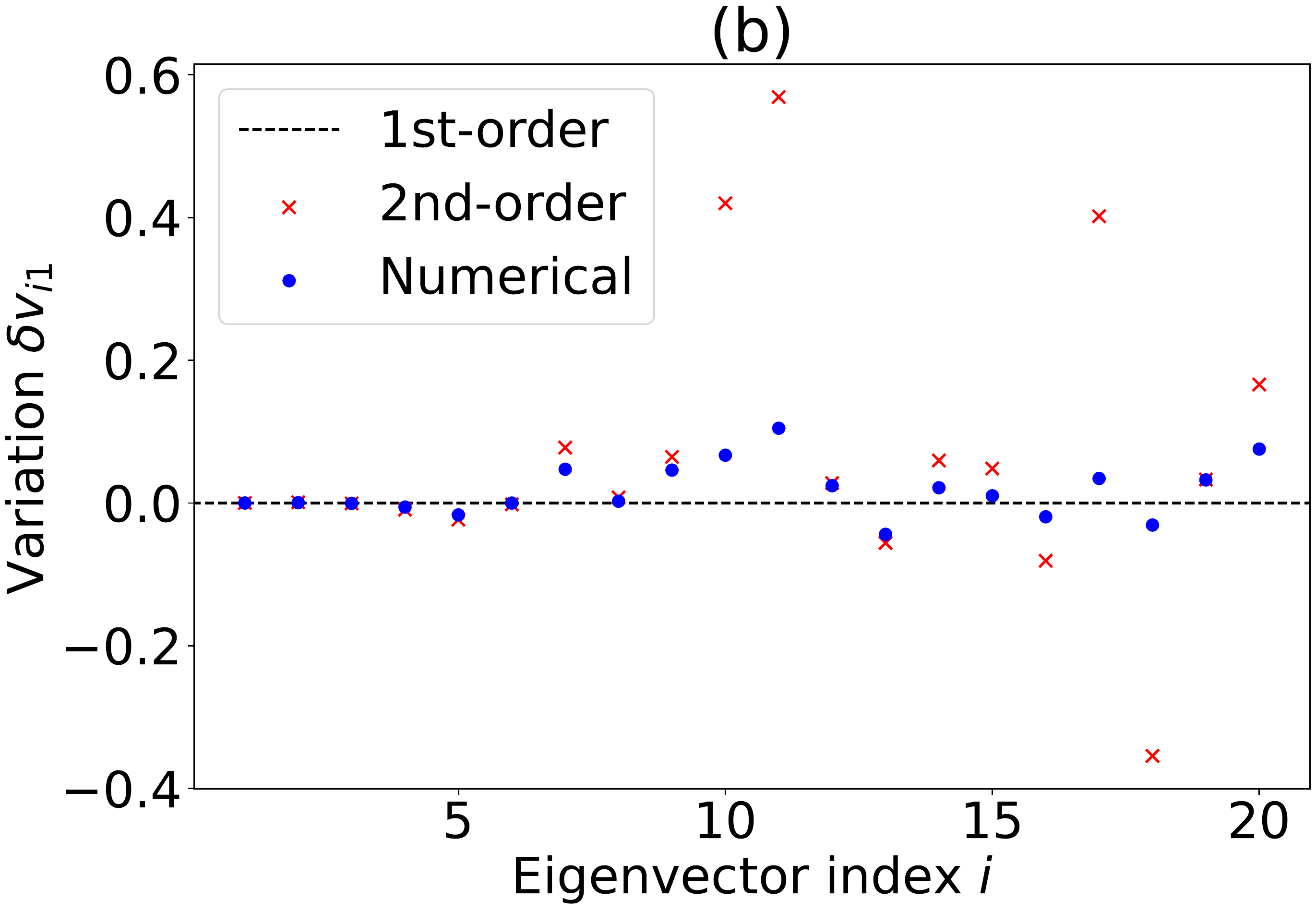}
    \end{minipage}
    \end{tabular}
    \caption{The second-order term for the variation of eigenvalues and eigenvector elements of the adjacency matrix for the undirectized macaque-cortex network under uniformly random directization with $q=0.1$. The blue data points for the numerical results and the dashed line for the first-order approximation are the same as those in Fig.~{\ref{fig-sec3-1}}. (a) Variation of the real part of the eigenvalues obtained from Eq.~(\ref{eq-app1-2ndeigval}) (black crosses). (b) Variation of the real part of the first element of the top 20 eigenvectors $v_{i1}$ obtained from Eq.~(\ref{eq-app1-2ndeigvec}) (black crosses).}
\label{fig-app1-1}
\end{figure*}

\section{\label{sec:app2}Spectrum out of the perturbative regime}

We here investigate the behavior of eigenvalues out of the perturbative regime using synthetic graphs generated from the stochastic block model (SBM) \cite{holland1983stochastic,karrer2011stochastic}, which is a random graph model with a preassigned module structure.
We particularly consider the SBM with two equally sized blocks, namely the symmetric SBM; we let $B_{1}$ and $B_{2}$ denote the vertex sets of the two blocks, with $|B_1|=|B_2|=N/2$.
For each pair of vertices, an edge is generated independently and randomly with probability 
$p_{\mathrm{in}}=2c_{\mathrm{in}}/N$ if the vertices belong to the same block. 
Otherwise, they are connected with probability $p_{\mathrm{out}}=2c_{\mathrm{out}}/N$. 
The average degree is given by $c=c_{\mathrm{in}}+c_{\mathrm{out}}$.

\begin{figure}[t!]
    \centering
    \includegraphics[width=8cm]{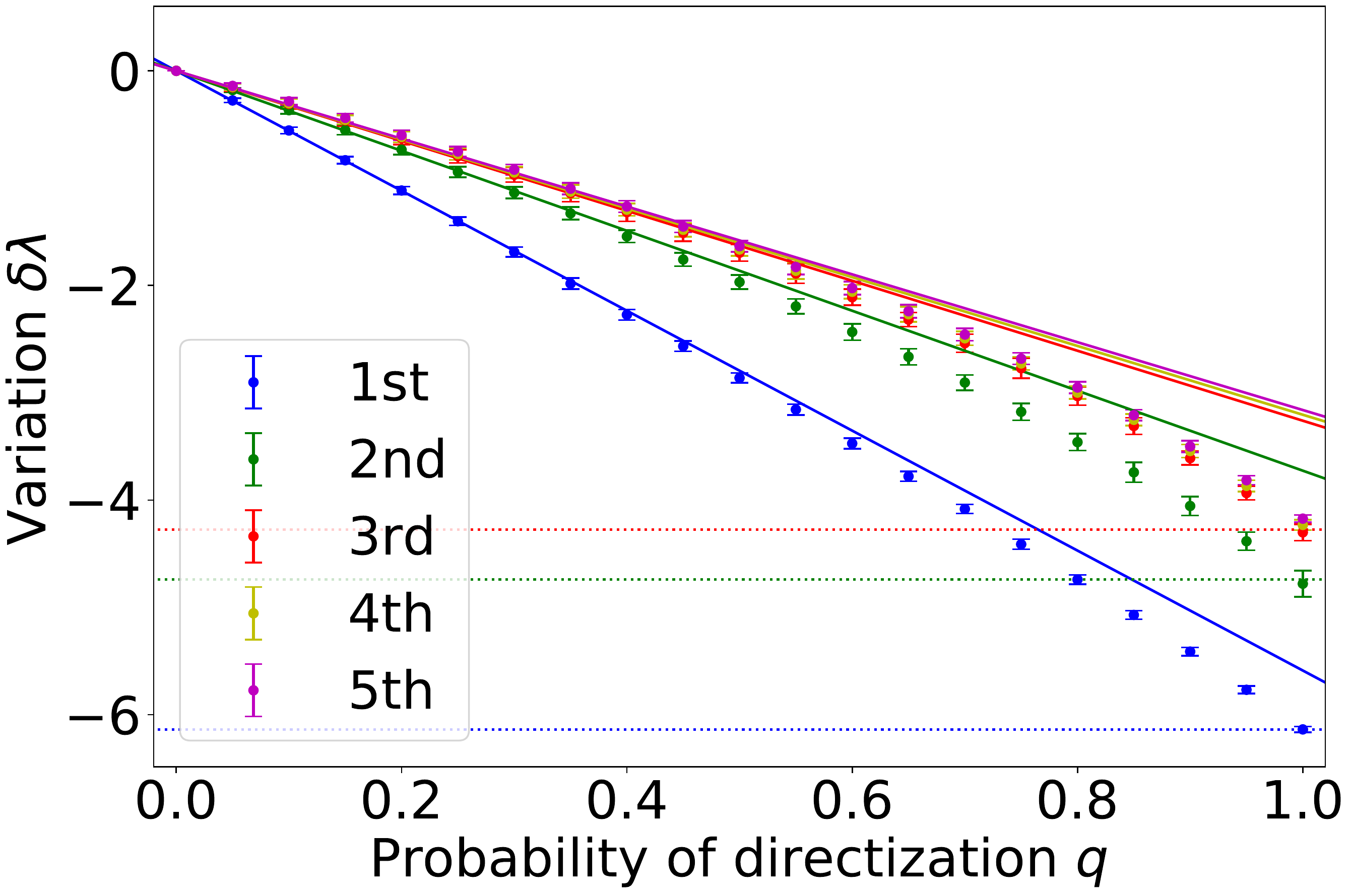}
    \caption{The variation of the top five eigenvalues along the real axis for an undirected graph generated by SBM with $N=1,000$, $c=10$ and $c_{\mathrm{out}}/c_{\mathrm{in}}=0.3$. Each point and error bar indicate the average and standard deviation over 100 directized samples, respectively. The solid lines indicate the theoretical estimates obtained from Eq.~(\ref{eq-sec3-eigval-avr}). The dotted lines represent the variation of eigenvalues for $q=1$ calculated using Eqs.~(\ref{eq-app3-edge}), (\ref{eq-app3-l1}) and (\ref{eq-app3-l2}).}
    \label{fig-app2-1}
\end{figure}
 
Figure~\ref{fig-app2-1} shows the variation $\delta\lambda$ of the top five eigenvalues for graphs generated from the SBM.
Similarly to the real-world networks in Fig.~\ref{fig-sec4}, it is confirmed that the perturbation theory is valid when $q$ is sufficiently small, and the differences between the estimate and numerical results increase as $q$ increases.

We can analytically estimate the variation of isolated eigenvalues when the graph is fully directized, i.e., $q=1$, by using the cavity method \cite{mezard1987spin,rogers2009cavity} for the symmetric SBM.
After full directization, the average number of in-neighbors in the same block and that in the different block are given by $c_\textrm{I}=c_\textrm{in}/2$ and $c_\textrm{O}=c_\textrm{out}/2$, respectively, while the average number of out-neighbors in the same block and that in the different block are also given by $c_\textrm{I}=c_\textrm{in}/2$ and $c_\textrm{O}=c_\textrm{out}/2$, respectively.
From the cavity method, the spectral band edge of the adjacency matrix spectrum in the $N\to\infty$ limit is given by \cite{neri2016eigenvalue,metz2019spectral,neri2020linear}
\begin{equation}
\lambda_{\mathrm{edge}}=\label{eq-app3-edge}\sqrt{2c}.
\end{equation}
Additionally, the cavity method yields the recursive equations, namely the cavity equations, for the eigenvector elements of the adjacency matrix. 
For a fully directed graph in the $N\to\infty$ limit, the right-eigenvector elements $\{r_i\}$ and the left-eigenvector elements $\{l_i\}$ corresponding to the eigenvalue $\lambda$ of the adjacency matrix are respectively given by the solution of the following cavity equation:
\begin{align}
r_j= \frac{1}{\lambda} \sum_{k\in \partial_j^{\mathrm{out}}} r_k,
\quad
l_j= \frac{1}{\lambda} \sum_{k\in \partial_j^{\mathrm{in}}} l_k,
\label{eq-app3-1}
\end{align}
where $\partial_j^{\mathrm{out}}$ and $\partial_j^{\mathrm{in}}$ represent the set of out-neighbors and that of in-neighbors, respectively \cite{neri2016eigenvalue,metz2019spectral}.

We use these equations to estimate the isolated eigenvalues for the symmetric SBM.
Considering the average over the symmetric SBM ensemble, we obtain the relations between the first and second moments of the eigenvector elements in each block as
\begin{align}
\langle r_{B_1} \rangle &=
\frac{1}{\lambda} \left(c_{\textrm{I}}\langle r_{B_1} \rangle + c_{\textrm{O}}\langle r_{B_2} \rangle \right), \\
\langle r_{B_2} \rangle &=
\frac{1}{\lambda} \left(c_{\textrm{O}}\langle r_{B_1} \rangle + c_{\textrm{I}}\langle r_{B_2} \rangle \right), \\
\langle r_{B_1}^2 \rangle &=
\frac{1}{\lambda^2}  
\left( c_{\textrm{I}} \langle r_{B_1}^2 \rangle + c_{\textrm{O}} \langle r_{B_2}^2 \rangle 
+c_{\textrm{I}}^2 \langle r_{B_1} \rangle^2 
+c_{\textrm{O}}^2 \langle r_{B_2} \rangle^2 
+2c_{\textrm{O}} c_{\textrm{I}}  \langle r_{B_1} \rangle \langle r_{B_2} \rangle \right), \\
\langle r_{B_2}^2 \rangle &=
\frac{1}{\lambda^2}  
\left( c_{\textrm{O}} \langle r_{B_1}^2 \rangle + c_{\textrm{I}} \langle r_{B_2}^2 \rangle 
+c_{\textrm{O}}^2 \langle r_{B_1} \rangle^2 
+c_{\textrm{I}}^2 \langle r_{B_2} \rangle^2 
+2c_{\textrm{O}} c_{\textrm{I}}  \langle r_{B_1} \rangle \langle r_{B_2} \rangle \right), \\
\langle l_{B_1} \rangle &=
\frac{1}{\lambda} \left(c_{\textrm{I}}\langle l_{B_1} \rangle + c_{\textrm{O}}\langle l_{B_2} \rangle \right), \\
\langle l_{B_2} \rangle &=
\frac{1}{\lambda} \left(c_{\textrm{O}}\langle l_{B_1} \rangle + c_{\textrm{I}}\langle l_{B_2} \rangle \right), \\
\langle l_{B_1}^2 \rangle &=
\frac{1}{\lambda^2}  
\left( c_{\textrm{I}} \langle l_{B_1}^2 \rangle + c_{\textrm{O}} \langle l_{B_2}^2 \rangle 
+c_{\textrm{I}}^2 \langle l_{B_1} \rangle^2 
+c_{\textrm{O}}^2 \langle l_{B_2} \rangle^2 
+2c_{\textrm{O}} c_{\textrm{I}}  \langle l_{B_1} \rangle \langle l_{B_2} \rangle \right), \\
\langle l_{B_2}^2 \rangle &=
\frac{1}{\lambda^2}  
\left( c_{\textrm{O}} \langle l_{B_1}^2 \rangle + c_{\textrm{I}} \langle l_{B_2}^2 \rangle 
+c_{\textrm{O}}^2 \langle l_{B_1} \rangle^2 
+c_{\textrm{I}}^2 \langle l_{B_2} \rangle^2 
+2c_{\textrm{O}} c_{\textrm{I}}  \langle l_1 \rangle \langle l_{B_2} \rangle \right), 
\end{align}
where the moments are taken over the SBM graph ensemble, $\langle r_{B_1}^k \rangle$ and $\langle l_{B_1}^k \rangle$ denote the $k$th moments of the former $|B_1|$ eigenvector elements, and $\langle r_{B_2}^k \rangle$ and $\langle l_{B_2}^k \rangle$ denote the $k$th moments of the latter $|B_2|$ eigenvector elements.
For isolated eigenvalues,  the first moments $\langle r_1 \rangle$, $\langle r_2 \rangle$, $\langle l_1 \rangle$ and $\langle l_2 \rangle$ are non-zero, which is satisfied when
\begin{equation}
\det \left( \begin{array}{cc}
c_\textrm{I}-\lambda & c_\textrm{O}\\
c_\textrm{O} & c_\textrm{I}-\lambda
\end{array} \right) =0.
\end{equation}
By solving this, we find the following two isolated eigenvalues for the symmetric SBM with two blocks:
\begin{align}
\lambda_1&=\label{eq-app3-l1} c_\textrm{I}+c_\textrm{O}=\frac{c}{2},\\
\lambda_2&=\label{eq-app3-l2} c_\textrm{I}-c_\textrm{O}=\frac{c_\textrm{in}-c_\textrm{out}}{2},
\end{align}
which are real.

The estimated variations for the spectral band edge and the two isolated eigenvalues obtained using Eqs.~(\ref{eq-app3-edge}), (\ref{eq-app3-l1}), and (\ref{eq-app3-l2}) are shown in Fig.~\ref{fig-app2-1} as the dashed lines.
Estimations from the cavity method and the numerical results coincide well.

\section{\label{sec:app4} Derivation of Eqs.~(\ref{eq-app1-2ndeigval}) and (\ref{eq-app1-2ndeigvec})}
\subsection{Eigenvalues}

From Eq.~(\ref{eq-app1-2ndeigval0}), we define the generating function for the second-order term of the $i$th eigenvalue $\delta\lambda_i^{(2)}$ by
\begin{equation}
Z_{ij}^{(2)}
= \left[
 e^{  \beta  \frac{\bm{v}^{\top}_i \bm{V} \bm{v}_j \bm{v}^{\top}_j \bm{V} \bm{v}_i}{\lambda_i-\lambda_j}  } 
  \right]_{\bm{V}|\bm{A}^0},
\end{equation}
which we transform as
\begin{align}
Z_{ij}^{(2)} 
&=\notag
\left[
\int du \int dw
\delta \left(u-\bm{v}^{\top}_i \bm{V} \bm{v}_j\right)
\delta \left(w-\bm{v}^{\top}_j \bm{V} \bm{v}_i\right)
e^{\frac{\beta u w}{\lambda_i-\lambda_j}}
 \right]_{\bm{V}|\bm{A}^0}\\
&=\notag
\left[
\int du \int dw \int \frac{ds}{2\pi} \int \frac{dt}{2\pi}
e^{is\left(u-\bm{v}^{\top}_i \bm{V} \bm{v}_j\right)}
e^{it\left(w-\bm{v}^{\top}_j \bm{V} \bm{v}_i\right)}
e^{\frac{\beta u w}{\lambda_i-\lambda_j}}
 \right]_{\bm{V}|\bm{A}^0}\\
&=\notag
\left[
\int du \int dw \int \frac{ds_j}{2\pi} \int \frac{dt_j}{2\pi}
e^{i\left(s u+t w\right)}
e^{\frac{\beta u w}{\lambda_i-\lambda_j}} 
\prod_{m,n}
e^{ -i  \left(s v_{im} v_{jn} + t v_{jm} v_{in}\right)  V_{mn}}  \right]_{\bm{V}|\bm{A}^0}\\
&= \notag
\int du \int dw \int \frac{ds}{2\pi} \int \frac{dt}{2\pi}
e^{i\left(s u+t w\right)}
e^{\frac{\beta u w}{\lambda_i-\lambda_j}} 
 \\& 
\quad
\mathop{\prod_{m<n}}_{\left(A^0_{mn}=1\right)}
\left(
 1-q 
+\frac{q}{2} e^{-i  \left(s v_{im} v_{jn} + t v_{jm} v_{in}\right) }
+\frac{q}{2} e^{-i  \left(s v_{in} v_{jm} + t v_{jn} v_{im}\right) }
\right).
\end{align}
Assuming that $q$ is small, we find
\begin{align}
Z_{ij}^{(2)} 
&\simeq\notag
\int du \int dw \int \frac{ds}{2\pi} \int \frac{dt}{2\pi}
e^{i\left(s u+t w\right)}
e^{\frac{\beta u w}{\lambda_i-\lambda_j}} 
 \\& 
\quad
\left(
 1-q \sum_{m<n} A^0_{mn}
 \left(1
-\frac{1}{2} e^{-i  \left(s v_{im} v_{jn} + t v_{jm} v_{in}\right) }
-\frac{1}{2} e^{-i  \left(s v_{in} v_{jm} + t v_{jn} v_{im}\right) }
\right) 
\right)\\
&=
\int du \int dw \int \frac{ds}{2\pi} \int \frac{dt}{2\pi}
e^{i\left(s u+t w\right)}
e^{\frac{\beta u w}{\lambda_i-\lambda_j}} 
\left(
 1-\frac{q}{2} \sum_{m,n} A^0_{mn}
 \left(1 - e^{-i  \left(s v_{im} v_{jn} + t v_{jm} v_{in}\right) } \right) 
\right)\\
&=
 1-\frac{q}{2} \sum_{m,n} A^0_{mn}
 \left(1 -  e^{\frac{\beta v_{im} v_{jn}  v_{jm} v_{in}}{\lambda_i-\lambda_j}} \right).
\end{align}
Thus, we obtain the average of the second-order term in the perturbation theory as
\begin{align}
\left[\delta\lambda_i^{(2)}\right]_{\bm{V}|\bm{A}^0}
&=\notag\sum_{j \neq i} \frac{\partial}{\partial \beta} \ln Z_{ij}^{(2)}|_{\beta=0}\\
&=\frac{q}{2} \sum_{j \neq i} \sum_{m,n} A^0_{mn} \frac{v_{im} v_{in} v_{jm} v_{jn} }{\lambda_i-\lambda_j},
\end{align}
which we used in Eq.~(\ref{eq-app1-2ndeigval}).

\subsection{Eigenvectors}

From Eq.~(\ref{eq-app1-2ndeigvec0}), we define the three generating functions for the second-order term of the $i$th eigenvector $\bm{v}_i$ as
\begin{equation}
x_{ij}^{(2)}=
\left[
 e^{  \beta \left(\frac{\bm{v}_j^{\top} \bm{V} \bm{v}_i}{\lambda_i-\lambda_j}\right)^2  }
\right]_{\bm{V}|\bm{A}^0}, \quad
y_{ijk}^{(2)}=
\left[ 
e^{ \beta \frac{\bm{v}^{\top}_j \bm{V} \bm{v}_k \bm{v}^{\top}_k \bm{V} \bm{v}_i}
{\left(\lambda_i -\lambda_j\right) \left(\lambda_i -\lambda_k\right)} }
\right]_{\bm{V}|\bm{A}^0}, \quad
z_{ij}^{(2)}=
\left[
 e^{ \beta \frac{\bm{v}^{\top}_j \bm{V} \bm{v}_i \bm{v}^{\top}_i \bm{V} \bm{v}_i}
{\left(\lambda_i -\lambda_j\right)^2} }
\right]_{\bm{V}|\bm{A}^0}.
\end{equation}
Then, the random average of the second-order term is given by
\begin{equation}
\left[\delta \bm{v}_i^{(2)}\right]_{V|A^0}=
-\frac{1}{2}
\sum_{j\neq i}
\left(\frac{\partial}{\partial \beta} \ln x_{ij}^{(2)}|_{\beta=0}\right)
\bm{v}_i
+
\sum_{j\neq i} \left(
\sum_{k\neq i} 
\left(\frac{\partial}{\partial \beta} \ln y_{ijk}^{(2)}|_{\beta=0}\right)
-
\left(\frac{\partial}{\partial \beta} \ln z_{ij}^{(2)}|_{\beta=0}\right)
\right)
\bm{v}_j.
\end{equation}
We find the first generating function in the form
\begin{align}
x_{ij}^{(2)}
&=\notag
\left[ 
\int du
\delta\left(u-\bm{v}^{\top}_j \bm{V} \bm{v}_i \right)
e^{\frac{\beta u^2}{\left(\lambda_i -\lambda_j\right)^2}}
\right] \\
&=\notag
\left[ 
\int du \int \frac{ds}{2\pi} 
e^{is\left(u-\bm{v}^{\top}_j \bm{V} \bm{v}_i \right)}
e^{\frac{\beta u^2}{\left(\lambda_i -\lambda_j\right)^2}}
\right]_{\bm{V}|\bm{A}^0}\\
&=\notag
\left[ 
\int du \int \frac{ds}{2\pi} 
e^{isu}
e^{\frac{\beta u^2}{\left(\lambda_i -\lambda_j\right)^2}}
\prod_{m,n}
e^{ -i s v_{jm} v_{in} V_{mn} }
\right]_{\bm{V}|\bm{A}^0}\\
&=
\int du \int \frac{ds}{2\pi} 
e^{isu}
e^{\frac{\beta u^2}{\left(\lambda_i -\lambda_j\right)^2}}
\mathop{\prod_{m<n}}_{\left(A^0_{mn}=1\right)}
\left(
1-q+ \frac{q}{2} e^{ -i s v_{jm} v_{in} } + \frac{q}{2} e^{-i -i s v_{jn} v_{im} }
\right).
\end{align}
For small $q$, we have
\begin{align}
x_{ij}^{(2)}
&\simeq\notag
\int du \int \frac{ds}{2\pi} 
e^{isu}
e^{\frac{\beta u^2}{\left(\lambda_i -\lambda_j\right)^2}}
\left(
1-q
\sum_{m<n} A_{mn}^0
\left(
1- \frac{1}{2} e^{ -i s v_{jm} v_{in} } - \frac{1}{2} e^{-i -i s v_{jn} v_{im} }
\right)
\right)\\
&=\notag
\int du \int \frac{ds}{2\pi} 
e^{isu}
e^{\frac{\beta u^2}{\left(\lambda_i -\lambda_j\right)^2}}
\left(
1-\frac{q}{2}
\sum_{m,n} A_{mn}^0
\left( 1- e^{ -i s v_{jm} v_{in} }  \right)
\right)\\
&=
1-\frac{q}{2}
\sum_{m,n} A_{mn}^0
\left( 1- e^{\frac{\beta v_{jm}^2 v_{in}^2}{\left(\lambda_i -\lambda_j\right)^2}} \right).
\end{align}
The second generating function gives
\begin{align}
y_{ijk}^{(2)}
&=\notag
\left[ 
\int du \int dw
\delta\left(u-\bm{v}^{\top}_j \bm{V} \bm{v}_k \right)
\delta\left(w-\bm{v}^{\top}_k \bm{V} \bm{v}_i \right)
e^{
 \frac{\beta u w}{\left(\lambda_i -\lambda_j\right) \left(\lambda_i -\lambda_k\right)}}
\right] \\
&=\notag
\left[ 
\int du \int dw
\int \frac{ds}{2\pi} \int \frac{dt}{2\pi}
e^{is \left(u-\bm{v}^{\top}_j \bm{V} \bm{v}_k \right)}
e^{it \left(w-\bm{v}^{\top}_k \bm{V} \bm{v}_i \right)}
e^{\frac{\beta u w}{\left(\lambda_i -\lambda_j\right) \left(\lambda_i -\lambda_k\right)}}
\right]_{\bm{V}|\bm{A}^0}\\
&=\notag
\left[ 
\int du \int dw
\int \frac{ds}{2\pi} \int \frac{dt}{2\pi}
e^{i\left(s u+t w\right)}
e^{\frac{\beta  u w}{\left(\lambda_i -\lambda_j\right) \left(\lambda_i -\lambda_k\right)} }
\prod_{m,n}
e^{ -i \left( s v_{jm} v_{kn} +t v_{km} v_{in} \right) V_{mn} }
\right]_{\bm{V}|\bm{A}^0}\\
&=\notag
\int du \int dw
\int \frac{ds}{2\pi} \int \frac{dt}{2\pi}
e^{i\left(s u+t w\right)}
e^{\frac{\beta u w}{\left(\lambda_i -\lambda_j\right) \left(\lambda_i -\lambda_k\right)} }
\\& \quad
\mathop{\prod_{m<n}}_{\left(A^0_{mn}=1\right)}
\left(
1-q+ \frac{q}{2} e^{ -i \left( s v_{jm} v_{kn} +t v_{km} v_{in} \right)} + \frac{q}{2} e^{-i \left( s v_{jn} v_{km} +t v_{kn} v_{im} \right)}
\right).
\end{align}
Again for small $q$, we have
\begin{align}
y_{ijk}^{(2)}
&\simeq\notag
\int du \int \frac{ds}{2\pi} 
e^{i\left(s u+t w\right)}
e^{\frac{\beta u w}{\left(\lambda_i -\lambda_j\right) \left(\lambda_i -\lambda_k\right)} }
\\&\notag \quad
\left(
1-q \sum_{m<n} A_{mn}^0
\left(
1- \frac{1}{2} e^{ -i \left( s v_{jm} v_{kn} +t v_{km} v_{in} \right)} - \frac{1}{2} e^{-i \left( s v_{jn} v_{km} +t v_{kn} v_{im} \right)}
\right)
\right)\\
&=\notag
\int du \int \frac{ds}{2\pi} 
e^{i\left(s u+t w\right)}
e^{\frac{\beta u w}{\left(\lambda_i -\lambda_j\right) \left(\lambda_i -\lambda_k\right)} }
\left(
1-\frac{q}{2} \sum_{m,n} A_{mn}^0
\left( 1- e^{ -i \left( s v_{jm} v_{kn} +t v_{km} v_{in} \right)}  \right)
\right)\\
&=
1-\frac{q}{2}
\sum_{m,n} A_{mn}^0
\left( 1-  e^{\frac{\beta v_{jm} v_{kn} v_{km} v_{in}}{\left(\lambda_i -\lambda_j\right) \left(\lambda_i -\lambda_k\right)} } \right).
\end{align}
Finally, the third generating function takes the form
\begin{align}
z_{i\ell}^{(2)}
&=\notag
\left[ 
\int dx \int dy
\delta\left(x-\bm{v}^{\top}_j \bm{V} \bm{v}_i \right)
\delta\left(y-\bm{v}^{\top}_i \bm{V} \bm{v}_i \right)
e^{ \frac{\beta x y}{\left(\lambda_i -\lambda_j\right)^2}}
\right]_{\bm{V}|\bm{A}^0}\\
&=\notag
\left[ 
\int dx \int dy
\int \frac{dp}{2\pi} \int \frac{dq}{2\pi}
e^{ip \left(x-\bm{v}^{\top}_j \bm{V} \bm{v}_i \right)}
e^{iq \left(y-\bm{v}^{\top}_i \bm{V} \bm{v}_i \right)}
e^{ \frac{\beta x y}{\left(\lambda_i -\lambda_j\right)^2} }
\right]_{\bm{V}|\bm{A}^0}\\
&=\notag
\left[ 
\int dx \int dy
\int \frac{dp}{2\pi} \int \frac{dq}{2\pi}
e^{i\left(p x+q y\right)}
e^{\frac{\beta  x y}{\left(\lambda_i -\lambda_j\right)^2}}
\prod_{m,n}
e^{-i\left(p v_{jm} v_{in} + q v_{im} v_{in} \right)V_{mn}}
\right]_{\bm{V}|\bm{A}^0}\\
&=\notag
\int dx \int dy
\int \frac{dp}{2\pi} \int \frac{dq}{2\pi}
e^{i\left(p x+q y\right)}
e^{\frac{\beta x y}{\left(\lambda_i -\lambda_j\right)^2}}
\\&
\quad
\mathop{\prod_{m<n}}_{\left(A^0_{mn}=1\right)}
\left(
1-q+
\frac{q}{2}e^{-i\left(p v_{jm} v_{in} + q v_{im} v_{in} \right)}
+
\frac{q}{2}e^{-i\left(p v_{jn} v_{in} + q v_{in} v_{in} \right)}
\right),
\end{align}
which is followed for small $q$ by
\begin{align}
z_{ij}^{(2)}
&\simeq\notag
\int du \int \frac{ds}{2\pi} 
e^{i\left(p x+q y\right)}
e^{\frac{\beta x y}{\left(\lambda_i -\lambda_j\right)^2}}
\\&\notag \quad
\left(
1-q \sum_{m<n} A_{mn}^0
\left(
1- 
\frac{1}{2}e^{-i\left(p v_{jm} v_{in} + q v_{im} v_{in} \right)}
-
\frac{1}{2}e^{-i\left(p v_{jn} v_{in} + q v_{in} v_{in} \right)}
\right)
\right)\\
&=\notag
\int du \int \frac{ds}{2\pi} 
e^{i\left(p x+q y\right)}
e^{\frac{\beta x y}{\left(\lambda_i -\lambda_j\right)^2}}
\left(
1-\frac{q}{2} \sum_{m,n} A_{mn}^0
\left( 1- e^{-i\left(p v_{jm} v_{in} + q v_{im} v_{in} \right)}  \right)
\right)\\
&=
1-\frac{q}{2}
\sum_{m,n} A_{mn}^0
\left( 1-  e^{\frac{\beta v_{jm} v_{in} v_{im} v_{in}}{\left(\lambda_i -\lambda_j\right)^2}} \right).
\end{align}
Thus, we arrive at the average of the second-order term in the perturbation theory as
\begin{align}
\left[\delta v_{i\ell}^{(2)}\right]_{V|A^0}=-\frac{q}{2} \sum_{j\neq i}
\sum_{m,n} A_{mn}^0  
\left(
\frac{1}{2} \left(\frac{v_{jm}v_{in}}{\lambda_i-\lambda_j}\right)^2 v_{i\ell}
+
\left(
-
\sum_{k\neq i} \frac{v_{jm} v_{kn} v_{km} v_{in}}{\left(\lambda_i -\lambda_j\right) \left(\lambda_i -\lambda_k\right)} 
+
\frac{v_{jm} v_{in} v_{im} v_{in}}{\left(\lambda_i -\lambda_j\right)^2}
\right) v_{j\ell}
\right),
\end{align}
which we used in Eq.~(\ref{eq-app1-2ndeigvec}).

\end{widetext}

\end{document}